\documentclass[twocolumn,prd,
 reprint,
nofootinbib,
 amsmath,amssymb,
 aps,
]{revtex4}

\usepackage{comment}
\usepackage{feynmp}
\usepackage[pdftex]{graphicx}
\usepackage{dcolumn}

\usepackage{color}
\usepackage{framed, color}
\definecolor{shadecolor}{rgb}{0.85,0.9,0.9} 
\usepackage{dsfont}
\usepackage{graphicx}
\usepackage{dcolumn}
\usepackage{bm}
\usepackage{rotating}
\usepackage{marvosym}
\usepackage{enumitem}
\usepackage{hyperref}

\begin{document}

\preprint{APS/123-QED}

\title{Signal inference with unknown response:\\ Calibration-uncertainty renormalized estimator}

\author{Sebastian Dorn\textsuperscript{1,2,}\footnote{sdorn@mpa-garching.mpg.de}, Torsten A. En{\ss}lin\textsuperscript{1,2}, Maksim Greiner\textsuperscript{1,2}, 
Marco Selig\textsuperscript{1,2}, and Vanessa Boehm\textsuperscript{1,2}}
\affiliation{\textsuperscript{1} Max-Planck-Institut f\"ur Astrophysik, Karl-Schwarzschild-Str.~1, D-85748 Garching, Germany\\
\textsuperscript{2} Ludwigs-Maximilians-Universit\"at M\"unchen, Geschwister-Scholl-Platz 1, D-80539 M\"unchen, Germany}
\date{\today}

\begin{abstract}
The calibration of a measurement device is crucial for every scientific experiment, where a signal has to be inferred from data. 
We present CURE, the calibration uncertainty renormalized estimator, to reconstruct a signal and simultaneously the instrument's calibration from the same 
data without knowing the exact calibration, but its covariance structure. 
The idea of CURE, developed in the framework of information field theory, is starting with an assumed calibration to successively include more and more portions of calibration uncertainty into the signal 
inference equations and to absorb the resulting corrections into renormalized signal (and calibration) solutions. Thereby, the signal inference and calibration
problem turns into solving a single system of ordinary differential equations and can be identified with common resummation techniques used in field theories.
We verify CURE by applying it to a simplistic toy example and compare it against existent self-calibration schemes, Wiener filter solutions,
and Markov Chain Monte Carlo sampling. We conclude that the method is able to keep up in accuracy with the best self-calibration methods and serves 
as a non-iterative alternative to it.


\end{abstract}

\maketitle

\section{Introduction}\label{sec:intro} 
\subsection{Motivation}
Data analysis is the link between theory and experiment, wherein a signal has to be inferred from measured data. For this purpose the transformation of a signal to
data, the measurement response, has to be understood precisely. The reconstruction of this response is called calibration. 

In the most simple case of a time independent instrument response, the calibration can be determined by measuring an a priori well known signal in 
a regime with neglectable noise level. This is commonly called external calibration. However, the assumption of time independency cannot be accepted in the majority
of cases. Of course the time dependency caused by, e.g., environmental factors, periodicities and systematics, 
or the signal itself, can be estimated with utmost effort. The resulting calibration, however, has still to be extrapolated into future time, where
the real measurement will be performed and where these influences will not be known exactly. What might be known, however, are their statistics. 
The resulting uncertainty consequently affects the signal reconstruction and has to be taken into account. 

There are methods, which improve the calibration by iteratively calibrating on a signal reconstruction and then improving the reconstruction 
using the new calibration. Such self-calibration (selfcal) schemes are widely in usage. They can, however, be prone to systematic biases since signal
and calibration are partly degenerate, i.e., a feature in the data could be caused by either of them and it is not guaranteed that the selfcal scheme 
does the correct choice automatically.

An improved selfcal scheme, which takes signal uncertainties in the calibration step into account was presented in Ref.~\cite{2013arXiv1312.1349E}.
Since also this new selfcal is an approximative solution to the complex inference problem, we ask if there is room for further improvement using information
field theory (IFT) \cite{2009PhRvD..80j5005E}. To this end we develop a calibration uncertainty renormalized estimator (CURE) for a signal, which incorporates calibration uncertainties
successively in a so-called renormalization flow equation. In comparison to existent approaches this method is non-iterative. 
For a review and discussion of previous work on existent calibration methods we point to Refs.\ \cite{2013arXiv1312.1349E,1991}.


\subsection{Structure of the work}
The remainder of this work is organized as follows. In Sec.\ \ref{sec:ift} we review the basics of the free and interacting IFT with focus on the latter.
Sec.\ \ref{sec:self-cal} represents the main part of the paper, where the calibration problem is introduced and CURE is derived.
The basic ideas as well as the main formulae of alternative selfcal schemes are also presented within this section. In Sec.\ \ref{sec:num} the performance of several signal
reconstruction methods is studied within a numerical toy example. Results are summarized in Sec.\ \ref{sec:remarks}.  

\section{Information field theory}\label{sec:ift}
To follow the derivation of an estimator with renormalized calibration uncertainty in the framework of IFT one has to be familiar with the concepts of interacting
IFT (see in particular Secs.\ \ref{sec:interacting}, \ref{sec:renorm}). Thus, a brief review might be helpful, but can be skipped by an advanced reader.
For this purpose we basically follow the Refs.\ \cite{2009PhRvD..80j5005E,2011PhRvD..83j5014E}, where a more detailed description of IFT
can be found.

\subsection{Basic formalism \& free theory}
Typically, a signal has to be inferred from data with the challenging question, how to do this in an optimal\footnote{Optimal with respect to,
e.g., minimizing the $\mathcal{L}^2$-error.} way? To reasonably answer this questions we first have to agree on a particular data model.

Within this work we assume that the data can be expressed by a discrete data tuple, $d=(d_1,\dots,d_m)^T\in\mathds{R}^m,~m\in\mathds{N}$, which is related to a signal $s$ by
\begin{equation}
d = Rs +n,
\end{equation}
where $R$ is a linear response operation acting on the signal, and $n=(n_1,\dots,n_m)^T\in\mathds{R}^m$ denotes some measurement noise. Contrary to data and noise,
the signal $s\equiv s(x),~x\in\mathcal{U}$ is considered to be a continuous quantity over some Riemannian manifold $\mathcal{U}$, i.e., a physical (scalar) field.
The linearity of the signal response, which transforms the continuous signal into data space, is valid for many physical measurements, e.g., observations of the 
cosmic microwave background and large scale structure in astronomy (cosmology), spectroscopy in different fields of physics, or medical imaging. 

We further assume the signal and noise to be uncorrelated, $\mathcal{P}(s,n)=\mathcal{P}(s)\mathcal{P}(n)$, and primarily Gaussian, i.e.,
$\mathcal{P}(s)=\mathcal{G}(s,S)$ and $\mathcal{P}(n)=\mathcal{G}(n,N)$ with related covariances 
$S=\left\langle ss^\dag\right\rangle_{(s|S)}$ and $N=\left\langle nn^\dag\right\rangle_{(n|N)}$, respectively. Here, we implicitly introduced the notation
\begin{equation}
\begin{split}
\mathcal{G}(a,A)\equiv &~ \frac{1}{\sqrt{|2\pi A|}}\exp\left(-\frac{1}{2}a^\dag A^{-1}a \right),~\text{and}\\
\left\langle~.~\right\rangle_{(a|A)} \equiv&~ \int\mathcal{D}a ~.~\mathcal{P}(a|A),
\end{split}
\end{equation}
where $\dag$ denotes a transposition and complex conjugation, $*$. The appropriate inner product of two fields $\{a,~b\}$ is defined by
$a^\dag b\equiv \int_{\mathcal{U}} d^{\dim\mathcal{U}} x~ a^*(x) b(x)$. If the conditions described above (known linear response, Gaussian signal and noise
with known covariances) are met, we term the theory a \textit{free theory}.

It is often convenient and common to focus on logarithmic probabilities by relating Bayes theorem \cite{Bayes01011763} to statistical physics,
\begin{equation}
\mathcal{P}(s|d)=\frac{\mathcal{P}(s,d)}{\mathcal{P}(d)} \equiv \frac{1}{\mathcal{Z}}\exp\left[-\mathcal{H}(s,d)\right].
\end{equation}
Here, we introduced the information Hamiltonian 
\begin{equation}
\mathcal{H}(s,d) \equiv -\ln\left[\mathcal{P}(s,d)\right],
\end{equation}
and the partition function
\begin{equation}
\mathcal{Z}(d) \equiv \mathcal{P}(d) = \int \mathcal{D}s \exp{\left[-\mathcal{H}(s,d)\right]}.
\end{equation}
Still considering the above free theory we find
\begin{equation}
\label{h_p_free}
\begin{split}
\mathcal{H}(s,d) =&~ \mathcal{H}_0 - j^\dag s + \frac{1}{2}s^\dag D^{-1}s,~\text{and}\\
\mathcal{Z}(d) =&~\sqrt{|2\pi D|}\exp\left(\frac{1}{2}j^\dag D j - \mathcal{H}_0 \right),
\end{split}
\end{equation} 
with the abbreviations
\begin{equation}
\label{abb}
\begin{split}
\mathcal{H}_0 =&~\frac{1}{2}\ln|2\pi N| +\frac{1}{2}\ln|2\pi S| +\frac{1}{2}d^\dag N^{-1}d,\\
D^{-1} =&~S^{-1} + R^\dag N^{-1}R,~\text{and}\\
j^\dag =&~d^\dag N^{-1} R,
\end{split}
\end{equation}
where the so-called information propagator, $D$, and the information source, $j$, have been introduced. $|~.~|$ denotes the determinant.

To exploit the whole machinery of statistical physics we additionally include a moment generating term, $J^\dag s$, into the partition function,
\begin{equation}
\mathcal{Z}(d,J) = \int \mathcal{D}s \exp{\left[-\mathcal{H}(s,d) +J^\dag s\right]}.
\end{equation}
The last definition permits to express the connected correlation functions (= cumulants) of a probability density function (PDF) via functional derivatives \cite{2009PhRvD..80j5005E},
\begin{equation}
\label{funcderiv}
\left\langle s(x_1)\dots s(x_n)\right\rangle^c_{(s|d)}\equiv \frac{\delta^n \ln\left[\mathcal{Z}(d,J)\right]}{\delta J(x_1)\dots\delta J(x_n)}\bigg\vert_{J=0}.
\end{equation}
Since we consider a Gaussian signal, its mean is equivalent to the well known Wiener filter \cite{wiener1964time} solution, 
\begin{equation}
\label{gen_wiener}
\left\langle s\right\rangle_{(s|d)}=Dj\equiv m_\mathrm{w}. 
\end{equation}
Its two point correlation function describes the uncertainty of the reconstruction, 
$\left\langle ss^\dag \right\rangle^c_{(s|d)}=\left\langle (s-m_\mathrm{w})(s-m_\mathrm{w})^\dag \right\rangle_{(s|d)}=D$ , and all cumulants
with $n>2$ vanish. Therefore, the posterior is Gaussian and given by
\begin{equation}
\mathcal{P}(s|d)=\mathcal{G}(s-m_\mathrm{w},D). 
\end{equation}

\subsection{n-th order perturbation theory}\label{sec:interacting}
Within the free theory we required the noise and in particular the signal to be Gaussian. However, this requirement cannot be met in some cases, e.g.,
in case noise or response are signal dependent, or simply a non-linear signal field. In the framework of IFT these scenarios can often\footnote{See Sec.\ \ref{sec:renorm} for cases in which such a treatment is not sufficient.} be described by a Taylor-expanded Hamiltonian \cite{2009PhRvD..80j5005E} composed of a free part, $\mathcal{H}_\text{free}$ (Eq.\ (\ref{h_p_free})), and a so-called interacting part, $\mathcal{H}_\text{int}$,
\begin{equation}
\label{hamint}
\mathcal{H} = \mathcal{H}_\text{free} + \underbrace{\sum_{n=0}^{\infty}\frac{1}{n!}\Lambda^{(n)}\left[s^{(n)} \right]}_{\equiv \mathcal{H}_\text{int}},
\end{equation}
where the deviation from Gaussianity is encoded in the anharmonic terms, $n>2$. The term $\Lambda^{(n)}\left[s^{(n)} \right]$ denotes a complete, 
fully symmetric\footnote{$\Lambda^{(n)}\equiv \frac{1}{n!}\sum_\pi \Lambda^{(n)}_{\pi(x_1,\dots,x_n)}$, with $\pi$ representing every permutation of $\{1,\dots,n \}$.}, 
contraction between the rank-$n$ tensor $\Lambda^{(n)}$ and
the $n$ fields $s^{(n)}=(s^1,\dots,s^n)$. If a decent estimate $m_0$ is known, one should Taylor-expand the Hamiltonian around this 
reference field $m_0$ in terms of residuals $\phi \equiv s-m_0$. A well working estimate is, for instance, the Wiener filter solution of the free theory, Eq.~(\ref{gen_wiener}). 
Using this reference field expansion often permits to truncate the Taylor-expansion earlier, since the anharmonic terms become smaller.

Analogously to the free theory we define the partition function,
\begin{equation}
\label{derive_part}
\begin{split}
\mathcal{Z}(d,J) =&~ \int \mathcal{D}s \exp{\left[-\mathcal{H}(s,d) +J^\dag s\right]}\\
		=&~ \int\mathcal{D}s \exp\left[ -\mathcal{H}_\text{int}\right]\exp{\left[-\mathcal{H}_\text{free} +J^\dag s\right]}\\
		=&~ \exp\left( -\mathcal{H}_\text{int}\left[\frac{\delta}{\delta J} \right]\right)\int\mathcal{D}s\exp{\left[-\mathcal{H}_\text{free} +J^\dag s\right]}\\
		\equiv &~\exp\left( -\mathcal{H}_\text{int}\left[\frac{\delta}{\delta J} \right]\right) \mathcal{Z}_\text{free}\\
		=&~\left(1 - \mathcal{H}_\text{int}\left[\frac{\delta}{\delta J} \right] + \frac{1}{2!}\mathcal{H}^2_\text{int}\left[\frac{\delta}{\delta J} \right]-\dots \right)\mathcal{Z}_\text{free}. 
\end{split}
\end{equation}
In principle, Eqs.\ (\ref{funcderiv}) and (\ref{derive_part}) enable to calculate all correlation functions of a PDF perturbatively. 
These calculations, however, are very uncomfortable and lengthy. Fortunately, there exists a well known diagrammatic treatment in analogy to 
quantum field theory and thermal field theory \cite{2009PhRvD..80j5005E}. 
E.g., including the first two correction terms, the signal mean $m$ is given by
\unitlength=1mm
\DeclareGraphicsRule{.1}{mps}{*}{}
\begin{equation}
\begin{split}
&m_x=~\parbox{12mm}{\begin{fmffile}{j_bare}
			\begin{fmfgraph}(10,10)
				\fmfleft{i}
				\fmfright{o}
				\fmf{plain}{i,o}
				\fmfdot{o}
			\end{fmfgraph}
	\end{fmffile}}+~\parbox{20mm}{\begin{fmffile}{j_1}
			\begin{fmfgraph}(10,10)
				\fmfleft{i}
				\fmfright{o}
				\fmf{plain}{i,o}
				\fmf{plain}{o,o}
				\fmfdot{o}
			\end{fmfgraph}
	\end{fmffile}}+~\parbox{20mm}{\begin{fmffile}{j_2}
			\begin{fmfgraph}(15,10)
				\fmfleft{i}
				\fmfright{o1,o2}
				\fmf{plain}{i,v,o1}
				\fmf{plain}{v,o2}
				\fmfdot{v,o2,o1}
			\end{fmfgraph}
	\end{fmffile}}+\dots\\
 &= D_{xy} \left(j_y -\frac{1}{2}\Lambda^{(3)}_{yzv}D_{zv} - \frac{1}{2}\Lambda^{(3)}_{yzv}(Dj)_z (Dj)_v\right) +\dots,
\end{split}
\end{equation}
where the ordering of diagrams corresponds to those of the equations and dots ($\dots$) representing the residual Feynman-series of correction terms.
The external dots ($\bullet$) represent source terms, internal dots vertices (the tensors $\Lambda^{(n)}$), and lines (\rule{1cm}{1pt})  propagator terms,
respectively. Repeated indices are to be integrated over.

The Feynman rules used in this work, which are necessary to switch between the mathematical expressions and the corresponding diagrams, 
can be found in App.\ \ref{sec:feyn}.

\subsection{Uncertainty renormalization}\label{sec:renorm}
\subsubsection{Motivation}\label{sec:motiv}
The approach of perturbative diagrammatic expansion is supposed to work well if the Hamiltonian is dominated by linear and quadratic terms.
That in turn means that the the tensors $\Lambda^{(n)}$ describing the deviation from Gaussianity are sufficiently small for the Feynman-series to converge.
This is, however, not always the case, e.g., within the calibration problem where the signal response cannot be known exactly due to
some potential time-dependencies or uncontrolled systematics. This calibration uncertainty can lead to large, non-vanishing terms $\Lambda^{(n)}$
as we show in Sec.\ \ref{sec:cure} of this paper.

Following the concept of Ref.\ \cite{2011PhRvD..83j5014E}, we can circumvent such a problem by including successively more and more small portions
of, e.g., calibration uncertainty into a signal inference equation. 
The basic idea is to include
only a sufficiently small amount of uncertainty per step to ensure the non-Gaussian (interaction) terms to be weak. Finally, this process results 
in a renormalized propagator, $\tilde{D}$, and information source, $\tilde{j}$. This process is called uncertainty renormalization \cite{2011PhRvD..83j5014E}.


 
\subsubsection{Concept}
For reasons of clarity and comprehensibility we skip the most general derivation and justification of uncertainty renormalization, which can be found in 
Ref.~\cite{2011PhRvD..83j5014E}, and focus more on the pragmatic procedure thereof. In the following we consider the Taylor-expanded, effective Hamiltonian
to be of the form of Eq.\ (\ref{hamint}). 
To suppress the strength of the non-Gaussian contributions we include a so-called expansion parameter, $\delta t \ll 1$, into the Hamiltonian, 
\begin{equation}
\label{hamintexp}
\mathcal{H} = \mathcal{H}_\text{free} + \delta t \sum_{n=0}^{\infty}\frac{1}{n!}\Lambda^{(n)}\left[s^{(n)} \right],
\end{equation}
and concentrate on this new Hamiltonian for a moment. For an appropriately small $\delta t$ the interaction terms become sufficiently small
and the diagrammatic expansion of Sec.\ \ref{sec:interacting} is justified again. Hence, by including the first correction terms into the propagator,
$D\rightarrow \tilde{D}_{\delta t}$, and into the information source, $j\rightarrow \tilde{j}_{\delta t}$, we obtain
\begin{equation}
\begin{split}
 \tilde{D}_{\delta t}=&~~\parbox{12mm}{\begin{fmffile}{prop_bare}
			\begin{fmfgraph}(10,10)
				\fmfleft{i}
				\fmfright{o}
				\fmf{plain}{i,o}
			\end{fmfgraph}
	\end{fmffile}}+~\delta t \bigg(~\parbox{20mm}{\begin{fmffile}{prop_1}
			\begin{fmfgraph}(15,10)
				\fmfleft{i}
				\fmfright{o}
				\fmf{plain}{i,v,o}
				\fmfdot{v}
			\end{fmfgraph}
	\end{fmffile}}\\
	&+~\parbox{20mm}{\begin{fmffile}{prop_2}
			\begin{fmfgraph}(15,10)
				\fmfleft{i}
				\fmfright{o1,o2}
				\fmf{plain}{i,v,o1}
				\fmf{plain}{v,o2}
				\fmfdot{v,o2}
			\end{fmfgraph}
	\end{fmffile}}+~\dots\bigg)+\mathcal{O}(\delta t^2),\\
	D\tilde{j}_{\delta t}=&~~\parbox{12mm}{\begin{fmffile}{j_bare}
			\begin{fmfgraph}(10,10)
				\fmfleft{i}
				\fmfright{o}
				\fmf{plain}{i,o}
				\fmfdot{o}
			\end{fmfgraph}
	\end{fmffile}}+~\delta t\bigg(\parbox{20mm}{\begin{fmffile}{j_1}
			\begin{fmfgraph}(10,10)
				\fmfleft{i}
				\fmfright{o}
				\fmf{plain}{i,o}
				\fmf{plain}{o,o}
				\fmfdot{o}
			\end{fmfgraph}
	\end{fmffile}}\\
	&+~\parbox{20mm}{\begin{fmffile}{j_2}
			\begin{fmfgraph}(15,10)
				\fmfleft{i}
				\fmfright{o1,o2}
				\fmf{plain}{i,v,o1}
				\fmf{plain}{v,o2}
				\fmfdot{v,o2,o1}
			\end{fmfgraph}
	\end{fmffile}} +~\dots\bigg) +\mathcal{O}(\delta t^2),
\end{split}
\end{equation} 
where the dots ($\dots$) represent all diagrams of order $\mathcal{O}(\delta t)$, i.e., all possible one-vertex diagrams. This way, $t\in [0,1]$ can be 
identified with a pseudo-time, which measures the accumulated uncertainty correction to the information propagator and source, and the expansion parameter
$\delta t$ represents the time step in which $D$ and $j$ are increased from their intermediate values, $D_t$ and $j_t$, to their one-step-renormalized
(but not final!) values $D_{t+\delta t}$ and $j_{t+\delta t}$, i.e. 
\begin{equation}
\begin{split}
&~D_t\rightarrow D_{t+\delta t},~\text{and}\\
&~j_t\rightarrow j_{t+\delta t}.
\end{split}
\end{equation}

We want to emphasize that $\delta t$ cannot simply be set to unity to obtain the fully renormalized propagator, $\tilde{D}$,
because this step would violate the justification of our perturbative expansion (see Sec.~\ref{sec:motiv}). However, a single step of this analytical \textit{resummation}
can be infinitesimally small, permitting for the formal definition of the derivatives \cite{2011PhRvD..83j5014E}
\begin{equation}
\begin{split}
\frac{dD_t}{dt}\equiv&~ \lim_{\delta t \rightarrow 0}\frac{D_{t+\delta t}-D_{t}}{\delta t}~\text{and}\\
\frac{dj_t}{dt}\equiv&~ \lim_{\delta t \rightarrow 0}\frac{j_{t+\delta t}-j_{t}}{\delta t},
\end{split}
\end{equation}
whereby the \textit{renormalization flow equations} can be formulated,
\begin{equation}
\label{cde2}
\begin{split}
\frac{dD_t}{dt}=&~~\parbox{20mm}{\begin{fmffile}{prop_1}
			\begin{fmfgraph}(15,10)
				\fmfleft{i}
				\fmfright{o}
				\fmf{plain}{i,v,o}
				\fmfdot{v}
			\end{fmfgraph}
	\end{fmffile}}+~\parbox{20mm}{\begin{fmffile}{prop_2}
			\begin{fmfgraph}(15,10)
				\fmfleft{i}
				\fmfright{o1,o2}
				\fmf{plain}{i,v,o1}
				\fmf{plain}{v,o2}
				\fmfdot{v,o2}
			\end{fmfgraph}
	\end{fmffile}}~+~\dots\\
D\frac{dj_t}{dt}=&~~\parbox{20mm}{\begin{fmffile}{j_1}
			\begin{fmfgraph}(10,10)
				\fmfleft{i}
				\fmfright{o}
				\fmf{plain}{i,o}
				\fmf{plain}{o,o}
				\fmfdot{o}
			\end{fmfgraph}
	\end{fmffile}}+~\parbox{20mm}{\begin{fmffile}{j_2}
			\begin{fmfgraph}(15,10)
				\fmfleft{i}
				\fmfright{o1,o2}
				\fmf{plain}{i,v,o1}
				\fmf{plain}{v,o2}
				\fmfdot{v,o2,o1}
			\end{fmfgraph}
	\end{fmffile}}~+~\dots,
\end{split}
\end{equation}
which is a system of coupled differential equations for operators with boundary values $D_{t=0}=D$ and $j_{t=0}=j$. By solving these equations
one obtains the fully renormalized quantities $\tilde{D}=D_{t=1}$, $\tilde{j}=j_{t=1}$, and the renormalized Wiener filter formula
\begin{equation}
\label{normwien}
\tilde{m} = \tilde{D}\tilde{j}.
\end{equation}
This means, by solving Eq.~(\ref{cde2}), we finally calculate a Gaussian approximation to the correct posterior mean 
of $s$, $P(s|d)\approx \mathcal{G}(s-\tilde{m},\tilde{D})$.
\section{Self-calibration}\label{sec:self-cal}
Now we address the calibration problem, i.e., how to infer a physical signal field given a data set without precise knowledge of the signal response.
We consider the case in which an external calibration is not possible (see Sec.\ \ref{sec:intro}). 
Thus, the instrument has to be self-calibrated during the measurement process. If we had absolutely no information about the signal response (how a measurement
device transforms the signal into data) there would be absolutely no chance to infer the signal appropriately. However, if we have some information about the 
statistics of the response, e.g., the two point correlation function, this task becomes solvable. For this purpose we introduce the CURE method in the framework 
of IFT (Sec.\ \ref{sec:cure}) and review already existing methods (Sec.\ \ref{sec:selfcal}) to compare it against. 

The aim is to calculate an optimal\footnote{Optimal in the sense of minimizing the $\mathcal{L}^2$-error.} estimator for the signal (or in general the 
moments $\left\langle s \dots s\right\rangle_{(s|d)}$) 
given the data without exact information of the calibration. A way to approach this challenge is to consider the unknown calibration as a nuisance parameter, i.e., to marginalize over the calibration when calculating the signal posterior,
\begin{equation}
\mathcal{P}(s|d) = \int \mathcal{D}\gamma~ \mathcal{P}(s,\gamma|d) =\underbrace{\int \mathcal{D}\gamma~ \mathcal{P}(d,\gamma|s)}_{\mathcal{P}(d|s)}~ \frac{\mathcal{P}(s)}{\mathcal{P}(d)},
\end{equation}
which involves the calculation of the calibration marginalized likelihood. To do so, we assume the response to be a linear 
function in the calibration coefficients $\gamma_a$ with Gaussian statistics, i.e.~$R^\gamma \approx R^0 +\sum_a\gamma_aR^a$. The assumption of Gaussianity is
appropriate as long as we have a priori no
information about higher moments of $\gamma$, $\left\langle \gamma_1\dots \gamma_n\right\rangle_{(\gamma)}$ with $n>2$. The linearity can be considered as a
first order approximation around $\gamma_0=0$ in $\gamma$, 
\begin{equation}
\begin{split}
R^\gamma =&~ R(\gamma_0) + \frac{\partial R(\gamma)}{\partial \gamma_a}\bigg\vert_{\gamma = \gamma_0}\left(\gamma - \gamma_0 \right) + \mathcal{O}(\gamma^2)\\
	= &~R^0 +\sum_a\gamma_aR^a + \mathcal{O}(\gamma^2). 
\end{split}
\end{equation}
Under these assumptions one obtains \cite{2013arXiv1312.1349E,2002MNRAS.335.1193B}
\begin{equation}
\label{likeli}
\begin{split}
\mathcal{P}&(d|s) = \int \mathcal{D}\gamma~\mathcal{P}(d|s,\gamma)\mathcal{P}(\gamma)\\
 	=&~ \int \mathcal{D}\gamma~\mathcal{G}\left(d-\left(R^0 + \sum_a \gamma_a R^a \right)s,N\right)\mathcal{G}\left(\gamma,\Gamma\right)\\
	=&~ \mathcal{G}\left(d-R^0s,N + \sum_{ab}\Gamma_{ab}R^a ss^\dag {R^b}^\dag\right).
\end{split}
\end{equation}

The data variance of this Gaussian likelihood, Eq.\ (\ref{likeli}), depends on the correlation structure of the calibration,
$\Gamma=\left\langle \gamma \gamma^\dag \right\rangle_{(\gamma|\Gamma)}$, as well as on the signal $s$.
This, in turn, results in a non-Gaussian posterior, $\mathcal{P}(s|d)\propto\mathcal{P}(d|s)\mathcal{P}(s)$, such that calculations of moments cannot be
done analytically anymore. In principle one can adapt posterior sampling techniques like Markov Chain Monte Carlo (MCMC) methods to calculate, e.g.,
the posterior mean, $m_\text{MCMC}$. These approaches, however, are usually very expensive, which increases the attractivity of developing (semi-)analytical methods.

\subsection{Calibration uncertainty renormalized estimator}\label{sec:cure}

Now, we apply the concept of uncertainty renormalization to the selfcal problem.
According to Sec.\ \ref{sec:renorm} we introduce an expansion parameter $\delta t\ll1$ in the ansatz:
\begin{equation}
\mathcal{P}(s|d)\propto \mathcal{G}\left(d-R^0s,N + \delta t\sum_{ab}\Gamma_{ab}R^a ss^\dag {R^b}^\dag\right)\mathcal{P}(s).
\end{equation} 
To simplify the notation we define an auxiliary para-meter $\Xi \equiv \sum_{ab}\Gamma_{ab}R^a ss^\dag {R^b}^\dag$
and assume a Gaussian signal prior, $\mathcal{P}(s)=\mathcal{G}\left(s-s_0,S\right)$, with the a priori mean $s_0\equiv\left\langle s \right\rangle_{(s)}$. 

The Hamiltonian becomes 
\begin{equation}
\label{hamintermed}
\begin{split}
\mathcal{H}(d,s) =&~-\ln \mathcal{P}(d,s)\\
	=&~ -\ln\left[ \mathcal{G}\left(d-R^0s,N +\delta t ~\Xi\right) \mathcal{G}\left(s-s_0,S\right) \right] \\
	=&~ \frac{1}{2}\ln\left|2\pi S\right| + \frac{1}{2}\ln\left|2\pi(N +\delta t ~\Xi)\right| \\
	 &~+\frac{1}{2}\left(d-R^0s\right)^\dag \left(N +\delta t ~\Xi\right)^{-1}\left(d-R^0s\right)\\
	&~ +\frac{1}{2}(s-s_0)^\dag S^{-1}(s-s_0).
\end{split}
\end{equation}
We can use that the expansion parameter $\delta t$ is small, i.e.\ $\delta t ~ \Xi \ll N$ 
(spectrally\footnote{Means that $\xi^\dag \delta t ~\Xi ~\xi \ll \xi^\dag N \xi ~\forall \xi \in \mathds{R}^m\backslash 0$.}), 
whereby the approximations
\begin{equation}
\label{tapprox}
\begin{split}
&\ln|2\pi (N+\delta t ~\Xi)|\approx \ln|2\pi N| +  \text{tr}\left(\delta t ~\Xi ~N^{-1}\right) ,~\text{and}\\
&\left(N + \delta t ~\Xi \right)^{-1}\approx N^{-1} - N^{-1}\delta t ~\Xi~ N^{-1}
\end{split}
\end{equation}
can be made. Using Eqs.\ (\ref{hamintermed}), (\ref{tapprox})  yields
\begin{equation}
\label{hamgen}
\mathcal{H}(d,s) =  \mathcal{H}^{\text{free}} +\delta t \sum_{n=2}^{4}\frac{1}{n!}\lambda^{(n)}\left[s^{(n)} \right]
\end{equation}
with

\begin{align}
\label{hamdefs}
\begin{split}
\mathcal{H}^{\text{free}} =&~\mathcal{H}_0 + \frac{1}{2}s^\dag D^{-1}s - j^\dag s,\\
\lambda^{(2)}[s,s] =&~\sum_{ab} \Gamma^{ab} \left(s^\dag M^{ba}s - {j^a}^\dag s s^\dag j^b   \right)\\
	&~ +1~\text{perm.},\\
\lambda^{(3)}[s,s,s] =&~\sum_{ab} \Gamma^{ab}\left(\frac{1}{2}{j^a}^\dag ss^\dag M^{b0}s + \text{cc}.\right)\\
	&~ +5~\text{perm.},\\
\lambda^{(4)}[s,s,s,s] =&~\sum_{ab} \Gamma^{ab}\left(-\frac{1}{2}s^\dag M^{0a}ss^\dag M^{b0}s\right)\\
	&~ +23~\text{perm.},
\end{split}
\end{align}
with permutations (perm.) with respect to $s$ and the abbreviations
\begin{equation}
\label{abb}
\begin{split}
\mathcal{H}_0 =&~\frac{1}{2}\ln|2\pi N| +\frac{1}{2}\ln|2\pi S| +\frac{1}{2}d^\dag N^{-1}d\\
	&~ + \frac{1}{2}s_0^\dag S^{-1}s_0\\
D^{-1} =&~\left(S^{-1} + {R^{0}}^\dag N^{-1}R^{0} \right),\\
j^\dag =&~d^\dag N^{-1} R^{0} + s_0^\dag S^{-1},\\
M^{ab} =&~{R^a}^\dag N^{-1} R^b,\\
{j^a}^\dag =&~d^\dag N^{-1} R^a.
\end{split}
\end{equation}
Terms higher than fourth order in the signal are dropped by making the approximation of Eq.~(\ref{tapprox}).

\subsubsection{Zero point expansion}
Since the information Hamiltonian, Eqs.\ (\ref{hamgen}), (\ref{hamdefs}), and (\ref{abb}), has the structure of Eq.\ (\ref{hamintexp}), we can start to derive
the renormalization flow equations. First, we consider (also for pedagogical reasons) the special case, in which the a priori signal mean is zero but the signal
two point statistic is known, i.e., we use a zero centered, Gaussian prior, $\mathcal{P}(s)=\mathcal{G}(s,S)$.

Following Sec.\ \ref{sec:renorm}, the interaction terms of Eq.\ (\ref{hamgen}) (Eq.\ (\ref{hamdefs})) 
can be absorbed in a so-called renormalized information propagator $\tilde{D}_{\delta t}$ and information source $\tilde{j}_{\delta t}$ of order $\delta t$. 
Including this (first) correction these quantities read

\newpage
\unitlength=1mm
\DeclareGraphicsRule{.1}{mps}{*}{}
\begin{widetext}
\begin{equation}
\label{zpexp}
\begin{split}
\left(\tilde{D}_{\delta t}\right)_{xy} =&~ D_{xy} + \delta t \left(- D_{xz}\lambda^{(2)}_{zv}D_{vy} - D_{xz}\lambda^{(3)}_{zvu}(Dj)_{v}D_{uy}
-\frac{1}{2}D_{xz}\lambda^{(4)}_{zvur}D_{vu}D_{ry} - \frac{1}{2}D_{xz}\lambda^{(4)}_{zvur}(Dj)_{v}(Dj)_{u}D_{ry}\right)\\
&~+ \mathcal{O}\left(\delta t^2\right)\\
=&~~\parbox{12mm}{\begin{fmffile}{prop_bare}
			\begin{fmfgraph}(10,10)
				\fmfleft{i}
				\fmfright{o}
				\fmf{plain}{i,o}
			\end{fmfgraph}
	\end{fmffile}}+~\delta t \bigg(~\parbox{20mm}{\begin{fmffile}{prop_1}
			\begin{fmfgraph}(15,10)
				\fmfleft{i}
				\fmfright{o}
				\fmf{plain}{i,v,o}
				\fmfdot{v}
			\end{fmfgraph}
	\end{fmffile}}+~\parbox{20mm}{\begin{fmffile}{prop_2}
			\begin{fmfgraph}(15,10)
				\fmfleft{i}
				\fmfright{o1,o2}
				\fmf{plain}{i,v,o1}
				\fmf{plain}{v,o2}
				\fmfdot{v,o2}
			\end{fmfgraph}
	\end{fmffile}}+~\parbox{20mm}{\begin{fmffile}{prop_3}
			\begin{fmfgraph}(15,10)
				\fmfleft{i}
				\fmfright{o}
				\fmf{plain}{i,v,o}
				\fmf{plain}{v,v}
				\fmfdot{v}
			\end{fmfgraph}
	\end{fmffile}}+~\parbox{20mm}{\begin{fmffile}{prop_4}
			\begin{fmfgraph}(15,10)
				\fmfleft{i}
				\fmfright{o}
				\fmftop{m1}
				\fmfbottom{m2}
				\fmf{plain}{i,v,o}
				\fmf{plain}{m1,v,m2}
				\fmfdot{v,m1,m2}
			\end{fmfgraph}
	\end{fmffile}}\bigg)+ \mathcal{O}\left(\delta t^2\right),\\
D_{xy}\left(\tilde{j}_{\delta t}\right)_y =&~D_{xy}\bigg[j_y +\delta t \bigg(- \frac{1}{2}\lambda^{(3)}_{yzv}D_{zv} -\lambda^{(2)}_{yz}(Dj)_{z} - \frac{1}{2}\lambda^{(3)}_{yzv}(Dj)_{z}(Dj)_{v} 
- \frac{1}{2}\lambda^{(4)}_{yzvu}D_{zv}(Dj)_{u}\\
&~ -\frac{1}{3!}\lambda^{(4)}_{yzvu}(Dj)_{z}(Dj)_{v}(Dj)_{u}\bigg)\bigg] + \mathcal{O}\left(\delta t^2\right)\\
=&~~\parbox{12mm}{\begin{fmffile}{j_bare}
			\begin{fmfgraph}(10,10)
				\fmfleft{i}
				\fmfright{o}
				\fmf{plain}{i,o}
				\fmfdot{o}
			\end{fmfgraph}
	\end{fmffile}}+~\delta t \bigg(~\parbox{20mm}{\begin{fmffile}{j_1}
			\begin{fmfgraph}(10,10)
				\fmfleft{i}
				\fmfright{o}
				\fmf{plain}{i,o}
				\fmf{plain}{o,o}
				\fmfdot{o}
			\end{fmfgraph}
	\end{fmffile}}+~\parbox{20mm}{\begin{fmffile}{j_5}
			\begin{fmfgraph}(15,10)
				\fmfleft{i}
				\fmfright{o}
				\fmf{plain}{i,v,o}
				\fmfdot{o,v}
			\end{fmfgraph}
	\end{fmffile}}+~\parbox{20mm}{\begin{fmffile}{j_2}
			\begin{fmfgraph}(15,10)
				\fmfleft{i}
				\fmfright{o1,o2}
				\fmf{plain}{i,v,o1}
				\fmf{plain}{v,o2}
				\fmfdot{v,o2,o1}
			\end{fmfgraph}
	\end{fmffile}}+~\parbox{20mm}{\begin{fmffile}{j_3}
			\begin{fmfgraph}(15,10)
				\fmfleft{i}
				\fmfright{o}
				\fmf{plain}{i,v,o}
				\fmf{plain}{v,v}
				\fmfdot{v,o}
			\end{fmfgraph}
	\end{fmffile}}+~\parbox{20mm}{\begin{fmffile}{j_4}
			\begin{fmfgraph}(15,10)
				\fmfleft{i}
				\fmfright{o}
				\fmftop{m1}
				\fmfbottom{m2}
				\fmf{plain}{i,v,o}
				\fmf{plain}{m1,v,m2}
				\fmfdot{v,m1,m2,o}
			\end{fmfgraph}
	\end{fmffile}}\bigg)\\
	&~+ \mathcal{O}\left(\delta t^2\right). 
\end{split}
\end{equation}
\end{widetext}

Just as a reminder, the vertices (internal dots) are multiplied by $\delta t$ while the source terms (external dots) are independent of $\delta t$. 
In the diagrammatic expansions, Eq.\ (\ref{zpexp}), we place $\delta t$ outside the brackets to underline this dependency. Therefore, to include all corrections
up to order $\delta t$, we have to include all possible one-vertex diagrams. It is crucial to realize that $\delta t$ cannot simply be set to one in order to
obtain the fully renormalized propagator, $\tilde{D}$, because this step would violate Eq.\ (\ref{tapprox}).
Appart from this it might also break down the perturbative expansion. However, instead of setting $\delta t=1$ we can interpret $t\in [0,1]$ as a pseudo-time,
which measures the accumulated correction to the information propagator and source (see Sec.\ \ref{sec:renorm}), $D_{t+\delta t}$ and $j_{t+\delta t}$. 
Thereby we can formulate the renormalization flow equations,
\begin{equation}
\label{cde}
\begin{split}
\frac{dD_t}{dt}=&~ \lim_{\delta t \rightarrow 0}\frac{D_{t+\delta t}-D_{t}}{\delta t}\\
		=&~~\parbox{20mm}{\begin{fmffile}{prop_1}
			\begin{fmfgraph}(15,10)
				\fmfleft{i}
				\fmfright{o}
				\fmf{plain}{i,v,o}
				\fmfdot{v}
			\end{fmfgraph}
	\end{fmffile}}+~\parbox{20mm}{\begin{fmffile}{prop_2}
			\begin{fmfgraph}(15,10)
				\fmfleft{i}
				\fmfright{o1,o2}
				\fmf{plain}{i,v,o1}
				\fmf{plain}{v,o2}
				\fmfdot{v,o2}
			\end{fmfgraph}
	\end{fmffile}}\\
			&~+~\parbox{20mm}{\begin{fmffile}{prop_3}
			\begin{fmfgraph}(15,10)
				\fmfleft{i}
				\fmfright{o}
				\fmf{plain}{i,v,o}
				\fmf{plain}{v,v}
				\fmfdot{v}
			\end{fmfgraph}
	\end{fmffile}}+~\parbox{20mm}{\begin{fmffile}{prop_4}
			\begin{fmfgraph}(15,10)
				\fmfleft{i}
				\fmfright{o}
				\fmftop{m1}
				\fmfbottom{m2}
				\fmf{plain}{i,v,o}
				\fmf{plain}{m1,v,m2}
				\fmfdot{v,m1,m2}
			\end{fmfgraph}
	\end{fmffile}},\\
D\frac{dj_t}{dt}=&~ D\left(\lim_{\delta t \rightarrow 0}\frac{j_{t+\delta t}-j_{t}}{\delta t}\right)\\
		=&~~\parbox{20mm}{\begin{fmffile}{j_1}
			\begin{fmfgraph}(10,10)
				\fmfleft{i}
				\fmfright{o}
				\fmf{plain}{i,o}
				\fmf{plain}{o,o}
				\fmfdot{o}
			\end{fmfgraph}
	\end{fmffile}}+~\parbox{20mm}{\begin{fmffile}{j_5}
			\begin{fmfgraph}(15,10)
				\fmfleft{i}
				\fmfright{o}
				\fmf{plain}{i,v,o}
				\fmfdot{o,v}
			\end{fmfgraph}
	\end{fmffile}}+~\parbox{20mm}{\begin{fmffile}{j_2}
			\begin{fmfgraph}(15,10)
				\fmfleft{i}
				\fmfright{o1,o2}
				\fmf{plain}{i,v,o1}
				\fmf{plain}{v,o2}
				\fmfdot{v,o2,o1}
			\end{fmfgraph}
	\end{fmffile}}\\
			&~+~\parbox{20mm}{\begin{fmffile}{j_3}
			\begin{fmfgraph}(15,10)
				\fmfleft{i}
				\fmfright{o}
				\fmf{plain}{i,v,o}
				\fmf{plain}{v,v}
				\fmfdot{v,o}
			\end{fmfgraph}
	\end{fmffile}}+~\parbox{20mm}{\begin{fmffile}{j_4}
			\begin{fmfgraph}(15,10)
				\fmfleft{i}
				\fmfright{o}
				\fmftop{m1}
				\fmfbottom{m2}
				\fmf{plain}{i,v,o}
				\fmf{plain}{m1,v,m2}
				\fmfdot{v,m1,m2,o}
			\end{fmfgraph}
	\end{fmffile}},
\end{split}
\end{equation}
which is a system of coupled differential equations for operators with boundary values $D_{t=0}=D$ and $j_{t=0}=j$. By solving these equations one obtains the fully
renormalized quantities $\tilde{D}=D_{t=1}$, $\tilde{j}=j_{t=1}$, and the renormalized Wiener filter formula
\begin{equation}
\label{normwien}
\tilde{m} = \tilde{D}\tilde{j}.
\end{equation}

However, instead of solving the coupled differential equations of Eq.\ (\ref{cde}) we could also solve the system where $dD_t/dt$ is replaced by an equivalently
valid equation for $dD^{-1}_t/dt$ leading to the new differential system
\begin{equation}
\label{renflow}
\begin{split}
\frac{dD^{-1}_{t,xy}}{dt}=&~\lambda^{(2)}_{xy} + \lambda^{(3)}_{xyz}(D_tj_t)_{z}+\frac{1}{2}\lambda^{(4)}_{xury}D_{t,ur}\\
&~ + \frac{1}{2}\lambda^{(4)}_{xvuy}(D_tj_t)_{v}(D_tj_t)_{u},\\
\frac{dj_{t,y}}{dt}= &~ - \frac{1}{2}\lambda^{(3)}_{yzv}D_{t,zv} - \frac{1}{2}\lambda^{(3)}_{yzv}(D_tj_t)_{z}(D_tj_t)_{v} \\
&~ -\lambda^{(2)}_{yz}(D_tj_t)_{z}
- \frac{1}{2}\lambda^{(4)}_{yzvu}D_{t,zv}(D_tj_t)_{u} \\
&~-\frac{1}{3!}\lambda^{(4)}_{yzvu}(D_tj_t)_{z}(D_tj_t)_{v}(D_tj_t)_{u}.
\end{split}
\end{equation}
Solving these equations might simplify the numerical effort in some cases. Afterwards we invert $D^{-1}_{t=1}\equiv \tilde{D}^{-1}$ to finally solve Eq.\ (\ref{normwien}). 

\subsubsection{Reference field expansion}
There is also the option to introduce a residual field $\phi = s - m_0$ with respect to a reference field, e.g., $m_0=Dj^0$ the Wiener filter solution without information of the proper 
calibration, Eq.~(\ref{gen_wiener}). By deriving a Hamiltonian of $\phi$ the perturbative expansion gets more exact while the non-Gaussian terms become smaller. 
The Hamiltonian then reads 
\begin{equation}
\label{hamgen_ref}
\mathcal{H}(d,\phi) =  \mathcal{H}_0' +\frac{1}{2}\phi^\dag D^{-1}\phi + \delta t \sum_{n=1}^{4}\frac{1}{n!}\Lambda^{(n)}\left[\phi^{(n)} \right],
\end{equation}
where $\mathcal{H}_0'$ includes all $\phi$-independent terms\footnote{Note that among the $\phi$-independent terms of $\mathcal{H}_0'$ are terms, collected in 
$\Lambda^{(0)}$, that depend on $\delta t$. These terms, however, only shift the Hamiltonian by a constant value but they do not influence its shape/structure.} 
and $\Lambda^{(n)}$ denotes the new (vertex-)tensor. Due to the fact that now already the source term is of $\mathcal{O}(\delta t)$ the diagrammatic expansion up
to order $\delta t$ reduces to a sum of Feynman diagrams containing only a single source and single vertex term, given by
\begin{equation}
\label{ref_expansion}
\begin{split}
&\left(\tilde{D}_{\delta t}\right)_{xy} = D_{xy}\\
&~+ \delta t \left( - D_{xz}\Lambda^{(2)}_{zv}D_{vy} -\frac{1}{2}D_{xz}\Lambda^{(4)}_{zvur}D_{vu}D_{ry}\right)\\
&=~~\parbox{12mm}{\begin{fmffile}{prop_bare}
			\begin{fmfgraph}(10,10)
				\fmfleft{i}
				\fmfright{o}
				\fmf{plain}{i,o}
			\end{fmfgraph}
	\end{fmffile}}+~\delta t \bigg(~\parbox{20mm}{\begin{fmffile}{prop_1}
			\begin{fmfgraph}(15,10)
				\fmfleft{i}
				\fmfright{o}
				\fmf{plain}{i,v,o}
				\fmfdot{v}
			\end{fmfgraph}
	\end{fmffile}}+~\parbox{20mm}{\begin{fmffile}{prop_3}
			\begin{fmfgraph}(15,10)
				\fmfleft{i}
				\fmfright{o}
				\fmf{plain}{i,v,o}
				\fmf{plain}{v,v}
				\fmfdot{v}
			\end{fmfgraph}
	\end{fmffile}}\bigg),\\
&~D_{xy}\left(\tilde{j}_{\delta t}\right)_y = D_{xy}\bigg[ \delta t \bigg(-\Lambda^{(1)}_{y} - \frac{1}{2}\Lambda^{(3)}_{yzv}D_{zv}\bigg)\bigg]\\
&=~\delta t \bigg(~\parbox{12mm}{\begin{fmffile}{j_bare}
			\begin{fmfgraph}(10,10)
				\fmfleft{i}
				\fmfright{o}
				\fmf{plain}{i,o}
				\fmfdot{o}
			\end{fmfgraph}
	\end{fmffile}}+~\parbox{20mm}{\begin{fmffile}{j_1}
			\begin{fmfgraph}(10,10)
				\fmfleft{i}
				\fmfright{o}
				\fmf{plain}{i,o}
				\fmf{plain}{o,o}
				\fmfdot{o}
			\end{fmfgraph}
	\end{fmffile}}\bigg). 
\end{split}
\end{equation}

After restoring the original signal $s$ by replacing the source term, $j_{\delta t}\rightarrow j_{\delta t} + D^{-1}_{\delta t}m_t,~m_t\equiv D_t j_t$ \cite{2013arXiv1312.1349E}, this
leads in analogy to the previous section to the renormalization flow equations,
\begin{equation}
\label{renflowref}
\begin{split}
\frac{dD_{t,xy}}{dt}=&~ - D_{xz}\Lambda^{(2)}_{zv}D_{vy} -\frac{1}{2}D_{xz}\Lambda^{(4)}_{zvur}D_{vu}D_{ry},\\
		\text{or}&~ \text{alternatively}\\
\frac{dD^{-1}_{t,xy}}{dt}=&~\Lambda^{(2)}_{xy} + \frac{1}{2}\Lambda^{(4)}_{xury}D_{t,ur},~\text{and}\\
\frac{dj_{t,y}}{dt}= &~ -\Lambda_y^{(1)} + \Lambda^{(2)}_{yz}(D_t j_t)_{z} \\
&~ - \frac{1}{2}\Lambda^{(3)}_{yzv}D_{t,zv} + \frac{1}{2}\Lambda^{(4)}_{yzvu}D_{t,zv}(D_tj_t)_{u},\\
\end{split}
\end{equation}
with boundaries $j_{t=0}=j^0$ and $D^{-1}_{t=0} = D^{-1}$. Note that the positive terms in the differential equation of $j_t$ arise from the
restoration of the original signal.

\begin{figure}[t]
\includegraphics[width=\columnwidth]{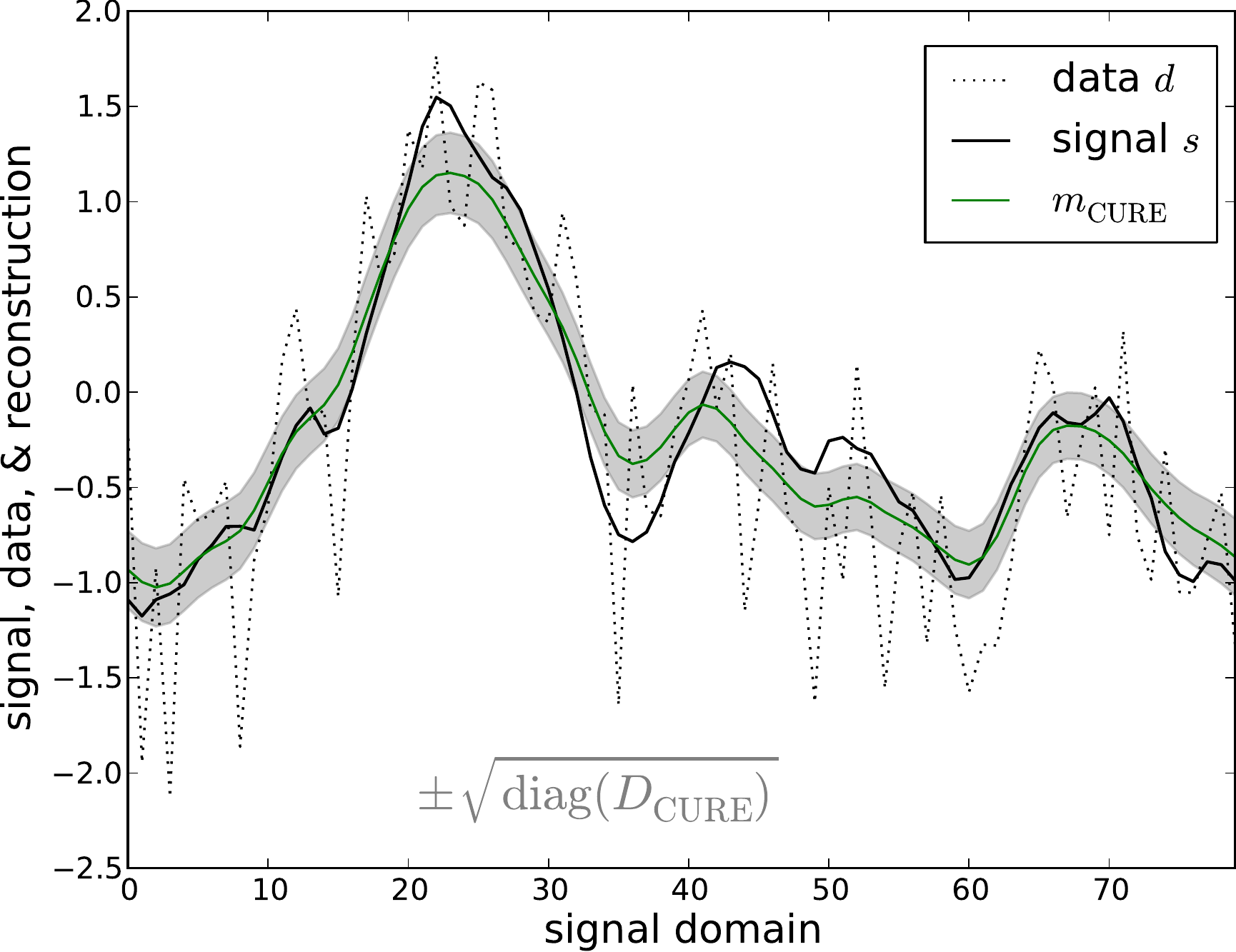}
\caption[width=\columnwidth]{(Color online) Signal, data, and signal reconstruction (considering unknown calibration) with related $1\sigma$-uncertainty according 
to Eq.\ (\ref{renflowref}) for the numerical example described in Sec.\ \ref{sec:num}. For comparison to other methods see 
Fig.~\ref{signal_recons}. Calibration and its reconstructions are shown in Fig.~\ref{gain_recons}.} 
\label{data_model}
\end{figure}

\begin{figure*}[t]
(a) \hspace{5.3cm} (b) \hspace{5.3cm} (c)\\
\includegraphics[width=.33\textwidth]{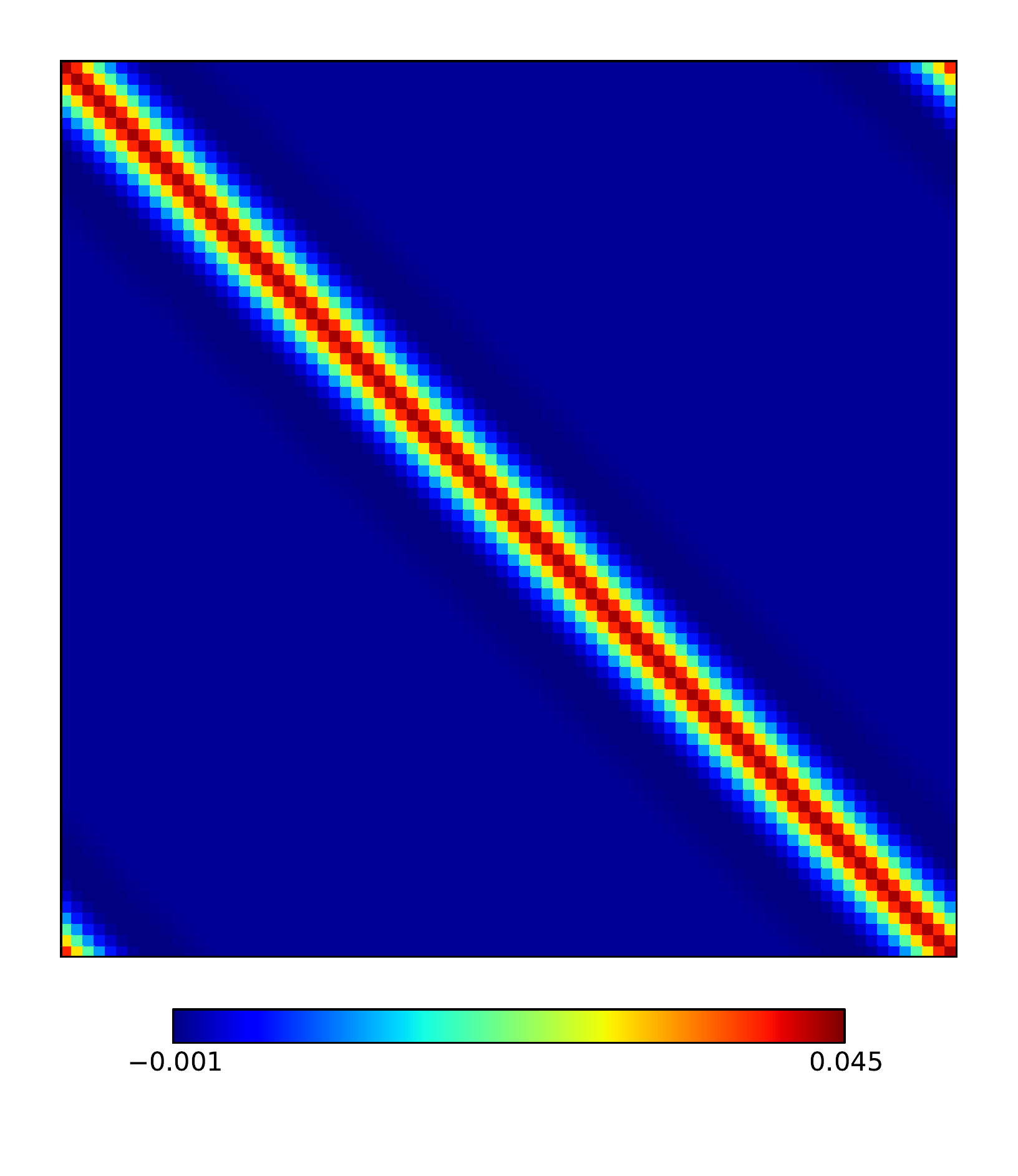}%
\includegraphics[width=.33\textwidth]{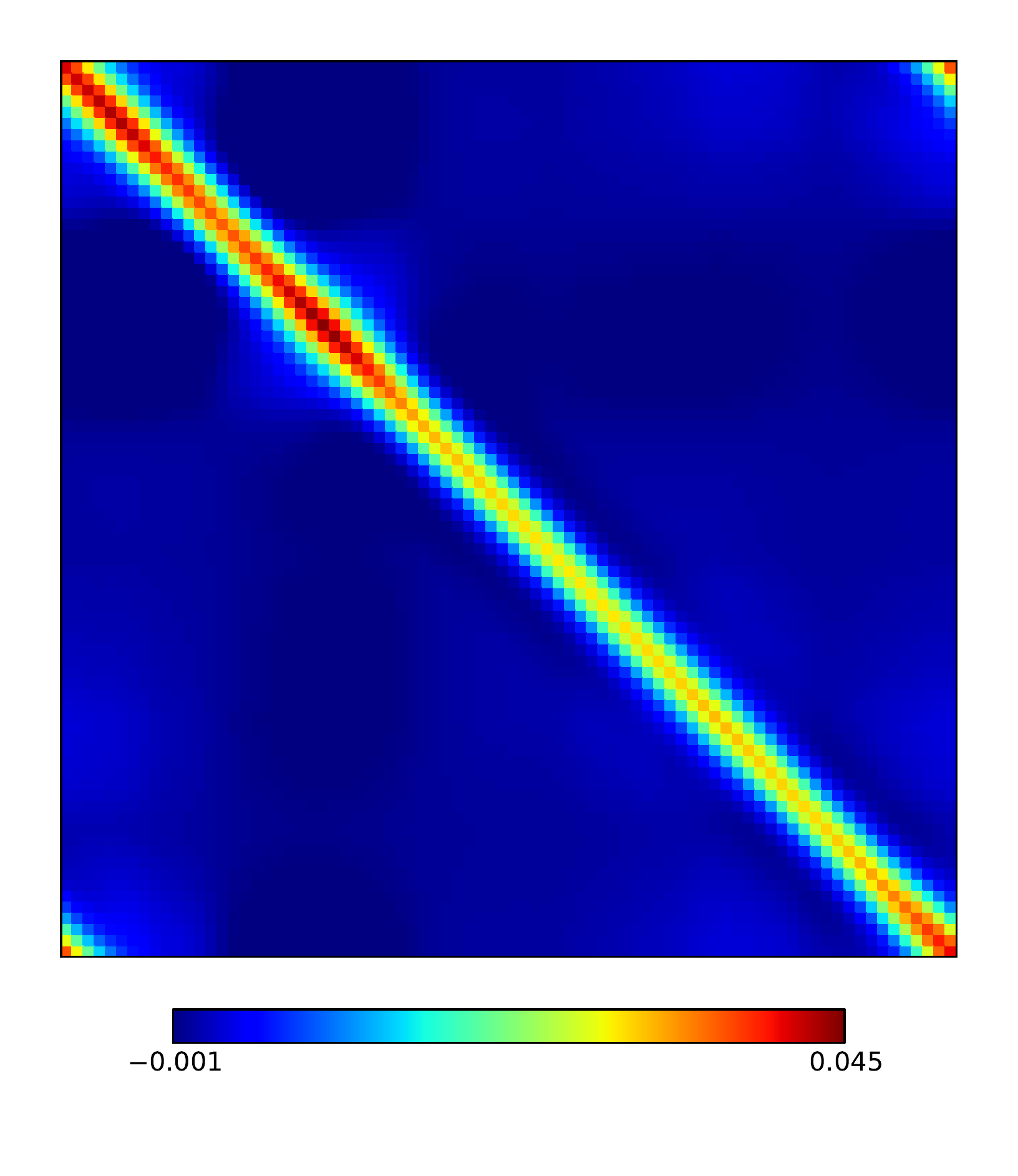}%
\includegraphics[width=.33\textwidth]{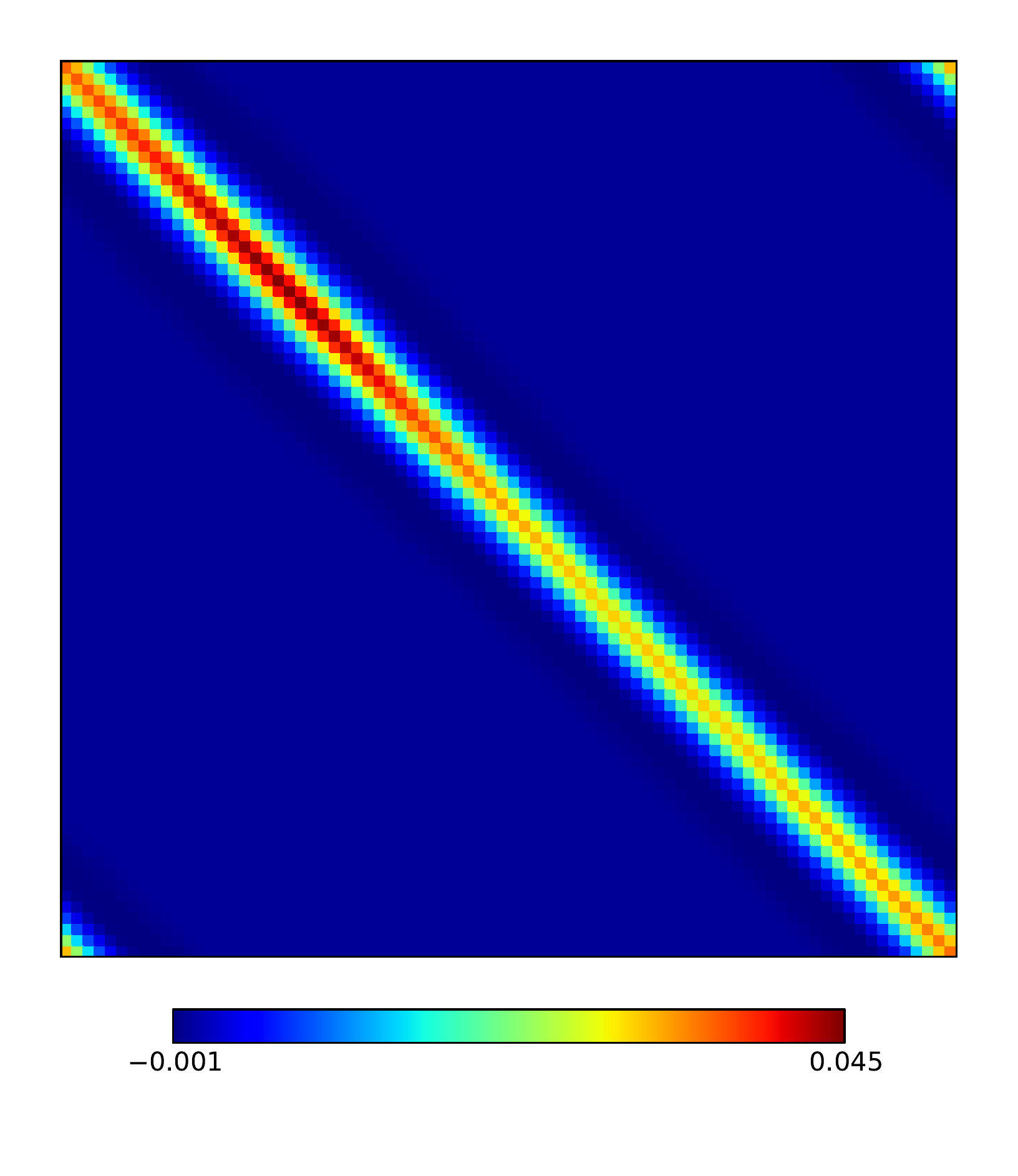}\\
(d) \hspace{5.3cm} (e) \hspace{5.3cm} (f)\\
\includegraphics[width=.33\textwidth]{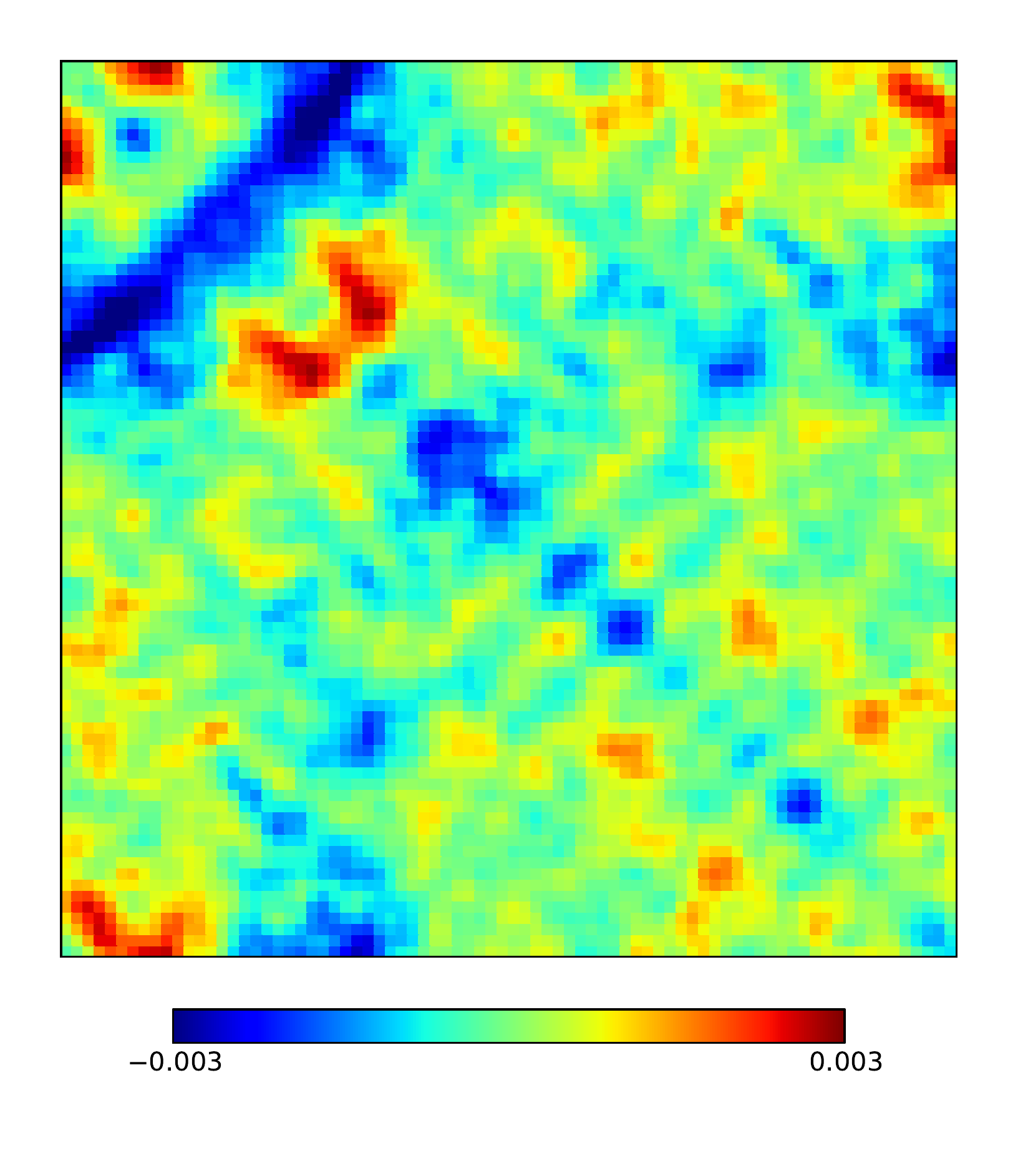}%
\includegraphics[width=.33\textwidth]{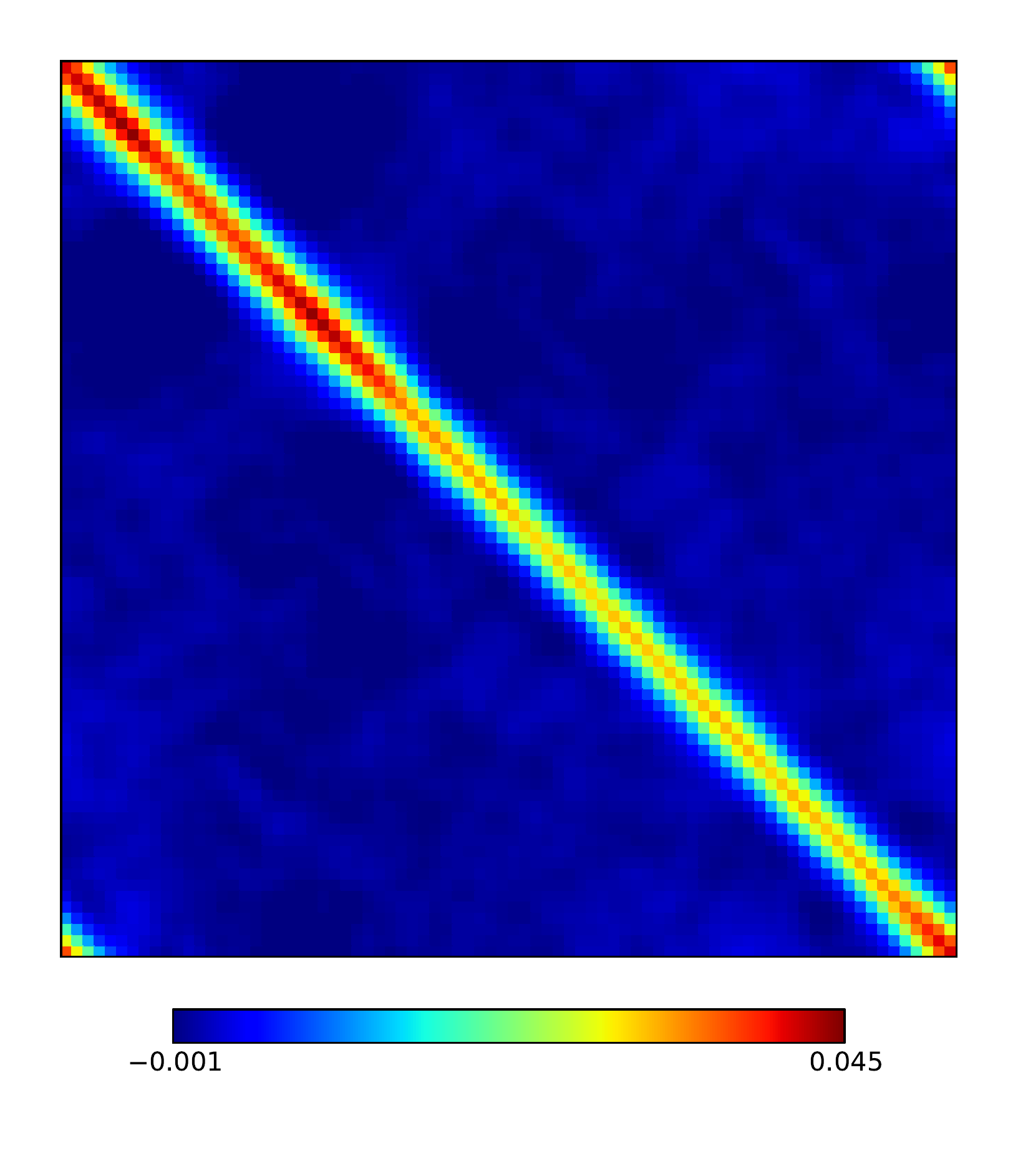}%
\includegraphics[width=.33\textwidth,height=.37\textwidth, clip=true,trim= 0mm -27mm 0mm 0mm]{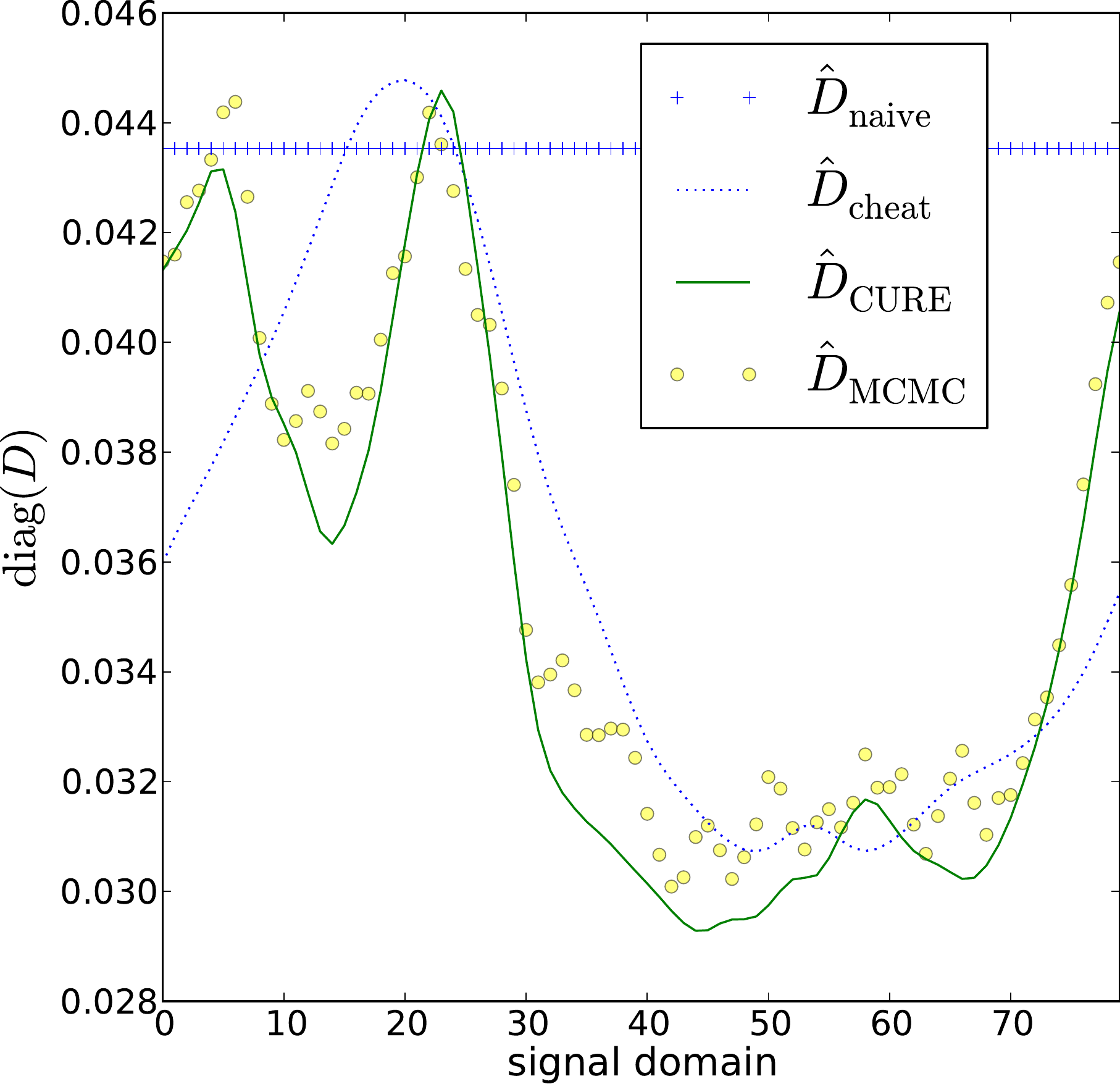}\vspace{-.5cm}%
\caption[width=\columnwidth]{\label{plot_prop}(Color online) Explicit structure of propagator operators for the realization shown in Fig.\ \ref{data_model},
\ref{signal_recons}, and \ref{gain_recons}. The figures (a)-(c), (e) refer to the propagators $D_\mathrm{method}$ with 
$\mathrm{method}=\mathrm{naive~(a),~CURE~(b),~cheat~(c),~and~MCMC~(e)}$ according to Eq.~(\ref{gen_wiener}) with unknown calibration
set to zero, Eq.~(\ref{renflowref}), and Eq.~(\ref{gen_wiener}) with known calibration, respectively. Upper panels \& lower middle panel: The renormalized propagator 
exhibits the same diagonal structure as the MCMC propagator. Lower panels (left, right): Comparison of the renormalized propagator to the MCMC result [CURE - MCMC, (d)] and explicit
structures of the propagator diagonals (denoted by $\hat{D}_\mathrm{method}$, (f)). Emerging from the process of resummation (involving absolute calibration  measurements, see Sec.~\ref{sec:num}),
Eq.\ (\ref{renflowref}), the renormalized propagator obtains a non-diagonal structure due to the complex, non-local vertex structure
of the non-Gaussian contributions to the Hamiltonian, see in particular Eq.\ (\ref{hamdefs_ref}).}  
\label{prop_plots}
\end{figure*}

Further note that the gained simplicity in the diagrammatic expansion has turned into a higher complexity of the vertex structure. 
The explicit structure of these vertices can be found in App.\ \ref{sec:app2}. These are also implemented for our numerical example, see Sec.\ \ref{sec:num} 
and Fig.\ \ref{data_model}. The effect of the resummation process (involving absolute calibration  measurements, see Sec.~\ref{sec:num}) on the information propagator is illustrated by Fig.~\ref{plot_prop}. 

\subsubsection{Approach optimization}\label{sec:opti}
Until now, the vertex-tensors $\Lambda^{(n)}$ were pseudo-time independent (see, for instance, Eq.\ (\ref{hamdefs_ref})). However, the CURE approach can
in principle be improved if we replace the residual field $\phi=s-m_0$ after every timestep by
$\tilde{\phi}_t =s-m_{t} = s-D_{ t}j_{ t}$. 
This way the support point of the expansion is always chosen optimally so that the first term of $\Lambda^{(1)}$ does still vanish (see Eq.~(\ref{hamdefs_ref}))
and the definition of time derivatives, Eq.\ (\ref{cde}), remains valid.

\subsection{Self-calibration schemes}\label{sec:selfcal}
To compare the derived CURE method not only against the Wiener filter solution without information of calibration (which is the starting value of CURE) but also to
two other iterative self-calibration (selfcal) schemes we review the basic ideas of the latter briefly. A full description of the following methods can
be found in Ref.\ \cite{2013arXiv1312.1349E}. The response is still considered to be linear, $R^\gamma = R^0 +\sum_a\gamma_aR^a$.

\subsubsection{Classical selfcal}
Classical selfcal is an iterative method, alternately inferring the signal while assuming the calibration to be known and vice versa until a fix-point is reached.
The respectively inferred quantities $s^\star$ and $\gamma^\star$ are often maximum a posterori (MAP) estimators. This procedure of simultaneously estimating $s$ and $\gamma$ can be identified
with searching for the maximum of the joint posterior, $\mathcal{P}(\gamma,s|d)$, or equally for the minimum of the joint information Hamiltonian \cite{2013arXiv1312.1349E},
$\mathcal{H}(d,\gamma,s) = -\ln [\mathcal{P}(d,\gamma,s)]$, given by
\begin{equation}
\frac{\partial \mathcal{H}(d,\gamma,s)}{\partial \gamma_a}\bigg\vert_{\gamma=\gamma^\star} = 0~\text{and}~
\frac{\partial \mathcal{H}(d,\gamma,s)}{\partial s}\bigg\vert_{s=s^\star} = 0.
\end{equation}
The resulting equations (Eq.\ (\ref{selfcal_gen}) with $T = 0$) must be iterated until convergence.

\subsubsection{New selfcal}
The new selfcal method is based on the above described idea of classical selfcal. However, in marked contrast to the latter new selfcal uses the signal marginalized
posterior to infer the calibration, and determines a signal estimate under the usage of the resulting calibration estimate and its uncertainty afterwards.
Therefore, the gradient and Hessian of the Hamiltonian, $\mathcal{H}(d,\gamma) = -\ln \int \mathcal{D}s ~\mathcal{P}(d,\gamma,s)$, have to be calculated to find 
the MAP estimate $\gamma^\star$ and its uncertainty $\Delta$, given by
\begin{equation}
\label{selfcal_der}
\frac{\partial \mathcal{H}(d,\gamma)}{\partial \gamma_a}\bigg\vert_{\gamma = \gamma^\star} = 0~\text{and}~
\frac{\partial^2 \mathcal{H}(d,\gamma)}{\partial \gamma_a \partial \gamma_b}\bigg\vert_{\gamma = \gamma^\star} \equiv \Delta^{-1}_{ab}.
\end{equation}
By following Ref.\ \cite{2013arXiv1312.1349E}, but skipping here the full derivation, we obtain the resulting calibration formula,

\begin{equation}
\label{selfcal_gen}
\begin{split}
\gamma^\star =&~ \Delta h,\\
\Delta_{ab}^{-1}=&~\Gamma^{-1}_{ab} + \text{tr}\left[\left( m m^\dag + TD\right)M^{ab} \right],~\text{and}\\
h_b =&~m^\dag j^b - \text{tr}\left[\left( m m^\dag + TD\right)M^{ab} \right],~\text{with}\\
T=&\left\{
    		\begin{array}{cc}
                		 1&~~~~~~~\text{for \text{new selfcal}}\\
                 		 0&~~~~~~~~~~\text{for \text{classic selfcal}}
    		\end{array} 
    		\right..
\end{split}
\end{equation} 
Note that the Wiener filter signal estimate $m = m(\gamma^\star)$ and its uncertainty $D=D(\gamma^\star)$ still depend on the calibration and thus $\Delta$ of Eq.\ (\ref{selfcal_gen}) is not exactly the one of Eq.\ (\ref{selfcal_der}) \cite{2013arXiv1312.1349E}.
For further details, as well as an extensive discussion of the selfcal methods we want to point to Ref.~\cite{2013arXiv1312.1349E}.

\begin{figure*}[htp]
\includegraphics[width=\textwidth]{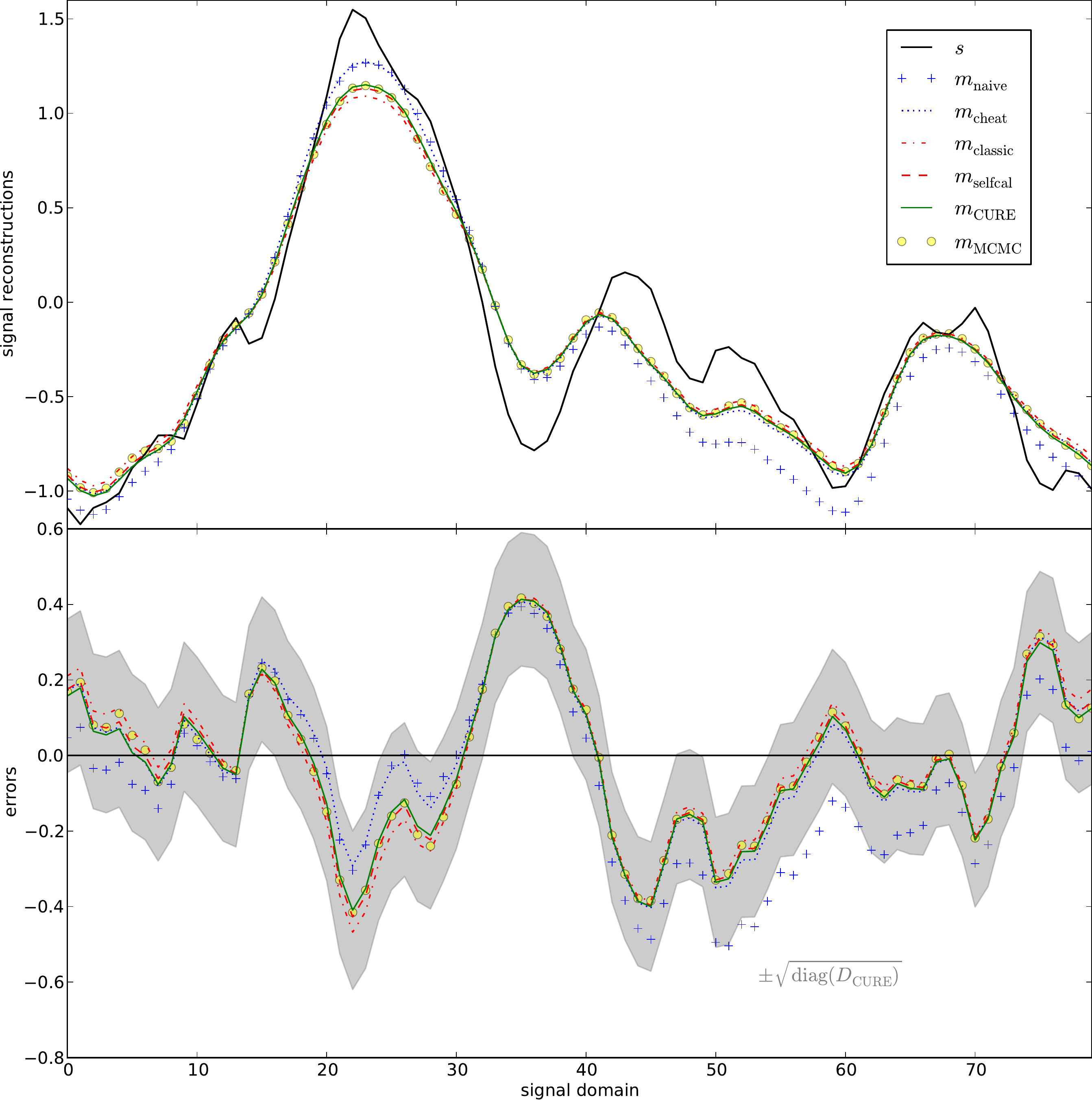}
\caption[width=\textwidth]{\label{signal_recons}(Color online) Signal reconstructions and related errors of different approaches. The following terminology is used: 
naive: Wiener filter with unknown calibration set to zero; classic: classical selfcal (Eq.\ (\ref{selfcal_abs}), $T=0$); CURE (Eqs.\ (\ref{renflowref}) and (\ref{hamdefs_ref}));
selfcal: new selfcal (Eq.\ (\ref{selfcal_abs}), $T=1$);
cheat: Wiener filter with known calibration; MCMC: Markov Chain Monte Carlo sampling. The
gray shaded region represents the $1\sigma$ uncertainty of the CURE method.} 
\end{figure*}

\section{Numerical Example}\label{sec:num}
\subsection{Setup \& results}
To demonstrate the efficiency of the derived CURE approach we address the illustrative, simplistic, one-dimensional example used in 
Ref.\ \cite{2013arXiv1312.1349E} and perform a direct comparison to the selfcal schemes and MCMC sampling (see, e.g., Ref.~\cite{1953JChPh..21.1087M}). 
There, a measurement device with a perfect point-like response scans a signal field $s$ over a (periodic)
domain $\Omega = \{x\}_x = [0,1)\subset \mathds{R}$ within a time $t\in[0,1)\subset \mathds{R}$, but with a time dependent calibration uncertainty,
given by the calibration coefficients $\gamma_t$. This instrument exhibits a sampling rate\footnote{Since this work is supposed to be a
proof of concept we work with explicit matrices and tensors, whereby we have to limit the size of the problem for computational reasons.
Further investigations are needed on how to transform this into a method using implicit tensors, and therefore suitable for ``big data'' problems.}
of $1/\tau = 80$ so that the $i$th data point,
measured at time $t=i\tau$, is related to the signal at position $x_t = i\tau$.  
During the measurement process spatial and temporal coordinates are aligned and the data are given by 
\begin{equation}
d_t = R_{tx}s_x +n_t = (1+\gamma_t)\delta(x-x_t)s_x +n_t,
\end{equation} 
where the signal, measurement noise $n_t$, and calibration coefficients $\gamma_t$ are Gaussian with $\mathcal{G}(s,S)$, $\mathcal{G}(n,N)$,
and $\mathcal{G}(\gamma,\Gamma)$ the corresponding PDF's with related covariance matrices $S,~N$, and $\Gamma$.
These are assumed to be known\footnote{In case they
are unknown there exist well known methods which are able to extract the correlation structure simultaneously from data, see, e.g., Ref.\ \cite{2011PhRvD..83j5014E}} and might
be described by their respective power spectra in Fourier space. Following Ref.\ \cite{2013arXiv1312.1349E} we use
\begin{equation}
\begin{split}
\label{power}
\mathcal{P}_s(k) =&~\frac{a_s}{\left(1+(k/k_s)^2\right)^{2}},\\
\mathcal{P}_\gamma(w) =&~\frac{a_\gamma}{\left(1+(w/w_\gamma)^2\right)^{2}},~\text{and}\\
\mathcal{P}_n(w) =&~ a_n.
\end{split}
\end{equation}
By Eq.\ (\ref{power}) the amplitudes $a_s = \sigma_s^2 \lambda_s$, $a_\gamma = \sigma_\gamma^2 \tau_\gamma$, and $a_n=\sigma_n^2 \tau_n$ with related
variances $\sigma^2_{s,\gamma,n}$ and correlation lengths $\lambda_s=4/k_s$, $\tau_\gamma = 4/\omega_\gamma$, and $\tau_n=\tau$ have been introduced. 
Within the numerical implementation we use the values $\sigma_s=1$, $\sigma_\gamma=0.3$, $\sigma_n=0.5$, $\lambda_s=0.3$ and $\tau_\gamma = 1.5$. 
This means we get an unit variance signal with calibration uncertainty of 30\% and noise of 50\%, which is still white
(percentage values with respect to the typical signal strength). 

Relating to Ref.\ \cite{2013arXiv1312.1349E} we also introduce so-called absolute calibration measurements to have additional information about the calibration
that is beneficial to break the global degeneracy of the data with respect to signal and calibration variations. This means, we switch off the signal for four
particular times $t_i\in\{0,0.25,0.5,0.75\}$, where the calibration has the strength $c=4$. Here, the data $d'$ is given by
\begin{equation}
d'_{t_i} = (1+\gamma_{t_i})c + n'_{t_i}.
\end{equation} 
During these measurements we assume the same noise statistics as before, $n' \hookleftarrow \mathcal{G}(n',N)$.

Including the absolute calibration measurements the iterative selfcal equations, Eq.\ (\ref{selfcal_gen}), become \cite{2013arXiv1312.1349E}
\begin{equation}
\label{selfcal_abs}
\begin{split}
\gamma^\star =&~ \Delta h,\\
\Delta_{tt'}^{-1}=&~\Gamma^{-1}_{tt'} + \sigma_n^{-2}\delta_{tt'}\left(q_t + c^2\sum_i \delta_{tt_i} \right),\\
h_t =&~\sigma_n^{-2}\left(d_t m_{x_t} - q_t + c^2\sum_i \delta_{tt_i}d_i' \right),~\text{and}\\
q_t =&~m_{x_t}^2 + TD_{x_tx_t}~\text{with}\\
T=&\left\{
    		\begin{array}{cc}
                		 1&~~~~~~~\text{for new selfcal}\\
                 		 0&~~~~~~~~~~\text{for classic selfcal}
    		\end{array} 
    		\right..
\end{split}
\end{equation}
To apply the CURE approach including the absolute calibration measurements we have to solve the 
ordinary differential equation of first order, according to Eq.\ (\ref{renflow}) or Eq.\ (\ref{renflowref}), depending on whether the \textit{zero-point} 
or \textit{reference field} expansion is used. We present here the more general, but more complex version of the \textit{reference field} expansion, Eqs.~(\ref{ref_expansion}) and (\ref{hamdefs_ref}), because 
this version is constructed to deal with a larger uncertainty of the calibration than the \textit{zero-point} expansion. 
To solve Eq.\ (\ref{renflowref}) we use the ordinary differential equation solver 
of {\tt scipy}\footnote{\url{http://docs.scipy.org/doc/scipy/reference/generated/scipy.integrate.ode.html}} with integrator 
settings: {\tt vode}, method = {\tt adams}. All numerical calculations have been performed using
\textsc{NIFTy}\footnote{\url{http://www.mpa-garching.mpg.de/ift/nifty}}~\cite{2013arXiv1301.4499S}. 

\begin{figure}[ht]
\includegraphics[width=\columnwidth]{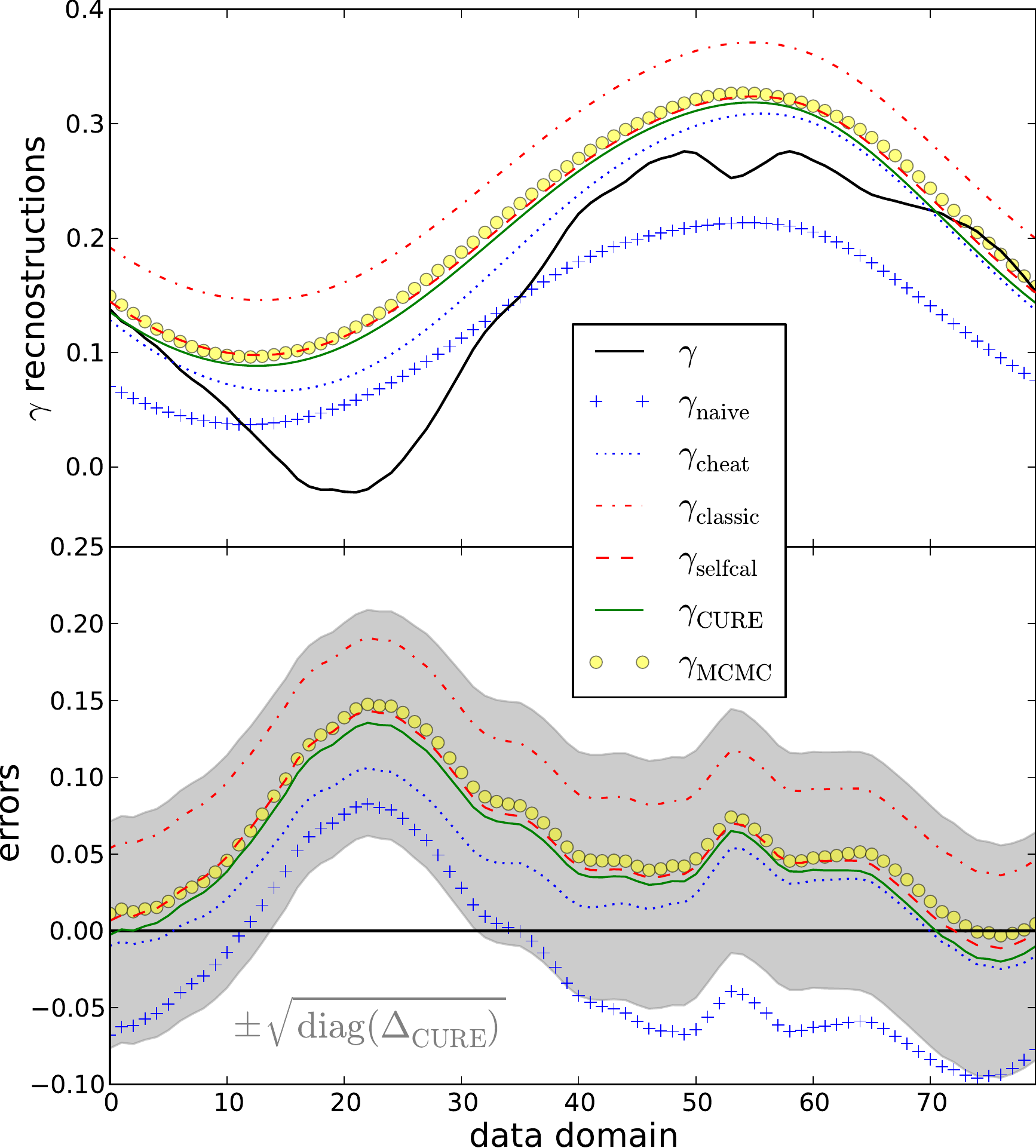}
\caption[width=\columnwidth]{\label{gain_recons}(Color online) Calibration reconstructions and related errors of different approaches using Eq.~(\ref{selfcal_abs}). 
The terminology is used  following Fig.\ \ref{signal_recons}. The gray shaded region represents the $1\sigma$ uncertainty of the CURE method. The reconstruction of 
cheat is not perfect, because Eq.~(\ref{selfcal_abs}) uses the Wiener filtered data (assuming the correct 
calibration, but non-neglectable noise). The relatively good result of naive is a pure coincidence.} 
\end{figure}

\begin{figure*}[htp]
\includegraphics[width=\textwidth]{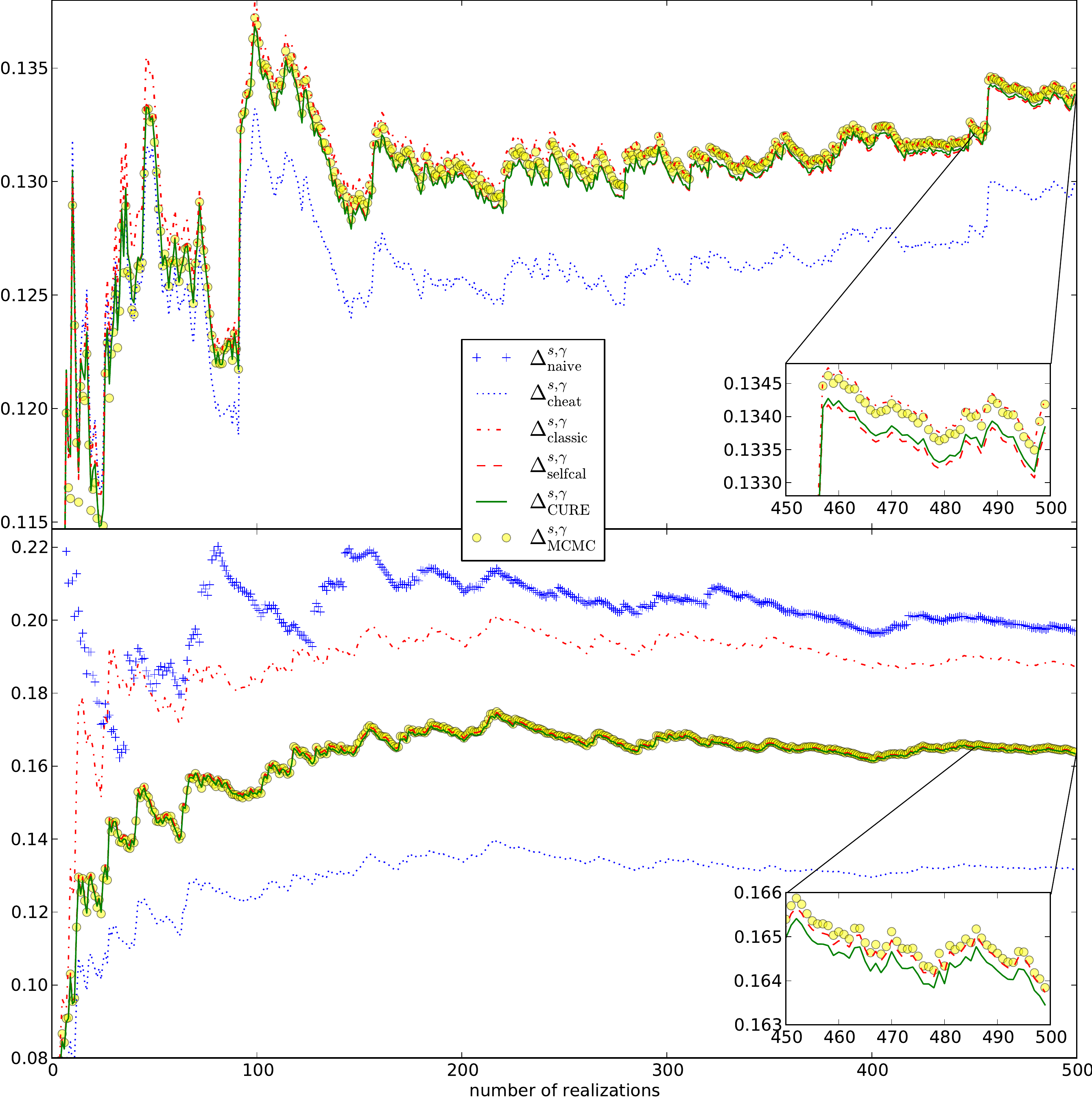}
\caption[width=\columnwidth]{(Color online) Squared error averages according to Eq.\ (\ref{error_stat}) at a given number of realizations for signal (upper panel) and calibration (lower panel). 
The best and worst result for signal and calibration yields the cheat and naive method, respectively. In the signal domain (upper panel) 
all three advanced methods are very close to each other, although there is a slight preference for the CURE and selfcal method followed by MCMC and
classic. The results of the naive method are beyond the range of the upper panel.
For the inference of calibration (lower panel) CURE and selfcal perform clearly better than classic and very similar to MCMC 
(see Tab.\ \ref{table1}, \ref{table2}, and \ref{table3}).} 
\label{statistics}
\end{figure*}

Figs.\ \ref{signal_recons} and \ref{gain_recons} show a typical result for signal and calibration 
reconstruction, respectively. 
Fig.\ \ref{statistics} and Tabs.~\ref{table1}, \ref{table2}, and \ref{table3} show the squared error averages of the different 
calibration methods according to Eq.\ (\ref{error_stat}) at a given number of realizations\footnote{Note that for the statistics of 500 realizations we use a four times coarser 
sampling rate.} for signal and calibration, where the following terminology is used,
\begin{equation}
\label{error_stat}
\begin{split}
\Delta_{i}^{s} \equiv &~ \left\langle (s - m_i)^\dag (s - m_i)\right\rangle_{(d,s,\gamma)},\\
\Delta_{i}^{\gamma} \equiv &~ \left\langle (\gamma - \gamma_i)^\dag (\gamma - \gamma_i)\right\rangle_{(d,s,\gamma)}~\text{with}\\
i=&~\text{naive},~\text{cheat},~\text{classic},~\text{selfcal},\\
&\text{CURE},~\text{and MCMC},
\end{split}
\end{equation}
referring to the Wiener filter methods without and with information of calibration, the classic and new selfcal scheme, the CURE scheme, 
and MCMC sampling, respectively.

\begin{table}[t]
\caption{\label{table1}
Squared errors of signal and calibration for all methods, averaged over 500 realizations, see Fig.~\ref{statistics}.}
\begin{ruledtabular}
\begin{tabular}{lcr}
$i$ &  $\Delta^{s}_i$ & $\Delta^{\gamma}_i$ \\
\colrule
naive & 0.1635& 0.1968\\
cheat & 0.1300 &  0.1316\\
classic & 0.1343 & 0.1873\\
selfcal& 0.1338 & 0.1637\\
CURE & 0.1338 & 0.1635\\
MCMC & 0.1342&0.1638\\
\end{tabular}
\end{ruledtabular}
\end{table}

\begin{table}[t]
\caption{\label{table2}
Improvements of the methods' signal squared errors with respect to the naive method, averaged over 500 realizations. }
\begin{ruledtabular}
\begin{tabular}{lcr}
$i$&
$\Delta^{s}_{\text{naive}} - \Delta^{s}_i$&
improvement\\
\colrule
naive & 0.0000  & 0.00\% \\
cheat&  0.0335& 100.00\% \\
classic & 0.0292 & 87.16\% \\
selfcal & 0.0297 & 88.86\% \\
CURE & 0.0297 & 88.66\% \\
MCMC & 0.0293&87.46\% \\
\end{tabular}
\end{ruledtabular}
\end{table}

\begin{table}[t]
\caption{\label{table3}
Improvements of the methods' calibration squared errors with respect to the naive method, averaged over 500 realizations.}
\begin{ruledtabular}
\begin{tabular}{lcr}
$i$&
$\Delta^{\gamma}_{\text{naive}} - \Delta^{\gamma}_i$&
improvement\\
\colrule
naive & 0.0000  & 0.00\% \\
cheat &  0.0652& 100.00\% \\
classic & 0.0095 & 14.57\% \\
selfcal& 0.0331 & 50.77\% \\
CURE & 0.0333 & 51.07\% \\
MCMC & 0.0330 & 50.61\% \\
\end{tabular}
\end{ruledtabular}
\end{table}
\subsection{Discussion}\label{sec:discuss}
As Figs.\ \ref{signal_recons}, \ref{gain_recons}, and in particular Fig.\ \ref{statistics} with related Tabs.~\ref{table1}, \ref{table2}, and \ref{table3} illustrate, 
the CURE and new selfcal (selfcal) approach prevail against classical selfcal (classic) and Wiener filtering with unknown calibration (naive)
and perform similar to the MCMC method. The latter represents in principle the best method by avoiding any approximations, but also the most expensive one.
Its small underperformance in comparison to CURE and selfcal has its origin in using still not sufficient samples for the MCMC-chains\footnote{For each signal
realization we have to run a separate chain. In the numerical example used in this work, a single chain consists of $2\times10^3$ independent samples.} to converge fully. Increasing their
number, however, would increase the numerical effort significantly.

The upside of CURE is that it is not iterative since it only involves the solution of a single system of coupled ordinary differential equations (ODE's).
For ODE's, in turn, exist a number of well working numerical solvers with adaptive
stepsize control, which might save significant amounts of computational time\footnote{We want to mention that we also found realizations for certain levels of 
signal, noise, and calibration where the {\tt scipy} solver did not converge. The reason for this might be a initial guess too far away from the correct solution.
In many cases one could cope with this problem by significantly reducing the stepsize, $\delta t$, or using the optimization described in Sec.~\ref{sec:opti}. We, however, did not elaborate on this since this work is supposed to be
a proof of concepts.}. This is, however, only true if one finds a clever 
implementation or sparse representation of $\Lambda^{(3)}$, because the 
term $\Lambda^{(3)}_{yzv}D_{zv}$, required in Eq.\ (\ref{ref_expansion}) might become a bottleneck within a calculation due to its complex correlation structure
(contrary to the $\Lambda^{(4)}$ term). 
Another downside is the higher level of complexity in comparison to new selfcal that naturally arises with a
renormalization calculation.

\section{Concluding remarks}\label{sec:remarks}
We derived the calibration uncertainty renormalized estimator (CURE) method to infer a signal and consequently the calibration without knowledge of the calibration but its covariance. 
The basic idea of CURE is to perform a perturbation calculation around a reference field, an a priori determined reconstruction of the signal without knowledge 
of calibration. Perturbatively means that we successively take into account higher-order terms of calibration uncertainty. 
This way, the problem of signal reconstruction without knowledge of the calibration, which is often solved by iterative or brute-force sampling methods, 
turns into a single system of ordinary differential equations.

We applied the method to a mock example and compared it against other existent calibration methods. For this example we found that CURE performs extremely similar 
to new~selfcal and MCMC sampling, and clearly beats the Wiener filter without calibration as well as the classic selfcal method in terms of reconstruction accuracy. 
Although it obviously performs well, a recommendation to favor this method over new and classical~selfcal depends on the particular problem at hand as well as on the numerical implementation,
as discussed in Sec.\ \ref{sec:discuss}. Therefore it serves as an alternative to them.

\begin{acknowledgments}
Calculations were realized using the \textsc{NIFTy} \cite{2013arXiv1301.4499S} package to be found at \url{https://github.com/mselig/nifty} as
well as {\tt scipy} to be found at \url{http://www.scipy.org}. 
\end{acknowledgments}

\newpage

\appendix
\begin{widetext}

\section{Feynman rules}\label{sec:feyn}
The Feynman rules originally stated in and inherited from Ref.\ \cite{2011PhRvD..83j5014E} read as follows:
\begin{enumerate}[noitemsep,topsep=2pt,parsep=2pt,partopsep=2pt]
\item Open ends of lines in diagrams correspond to external coordinates and are labeled by such. Since
the partition sum in particular does not depend on any external coordinate, it is calculated only
from summing up closed diagrams. However, the field expectation value $m(x) = \left\langle s(x)\right\rangle_{(s|d)}= \delta\ln[ \mathcal{Z}(d,J)]/\delta J(x)|_{J=0}$ and higher order correlation
functions depend on coordinates and therefore are calculated from diagrams with one or more open
ends, respectively.
\item A line with coordinates $x'$ and $y'$ at its end represents the propagator $D_{x'y'}$ connecting these locations.
\item Vertices with one leg get an individual internal,
integrated coordinate $x'$ and represent the term $j_{x'} +J_{x'} -\Lambda^{(1)}_{x'}$.
\item Vertices with $n$ legs represent the term $-\Lambda^{(n)}_{x'_1 \dots x'_n}$, where each individual leg is labeled by one of the
internal coordinates $x'_1 \dots x'_n$. This more complex vertex-structure, as compared to QFT, is a
consequence of non-locality in IFT.
\item All internal (and therefore repeatedly occurring)
coordinates are integrated over, whereas external
coordinates are not.
\item Every diagram is divided by its symmetry factor,
the number of permutations of vertex legs leaving
the topology invariant, as described in any book on
field theory.
\end{enumerate}

\section{Renormalization flow equations including absolute calibration measurements}\label{sec:app2}
This section derives the generalization of the renormalization flow equations in presence of absolute calibration measurements. These measurements can be included in the prior knowledge of the
calibration coefficients, $\mathcal{P}(\gamma) = \mathcal{G}(\gamma - m_\gamma, D^{\gamma})$, with $(m_\gamma)_a$ the Wiener filter solution for $\gamma$ with uncertainty $D^\gamma$
using the absolute calibration measurements only. Hence, the likelihood becomes
\begin{equation}
\label{likeliA}
\begin{split}
\mathcal{P}(d|s) &= \int \mathcal{D}\gamma~\mathcal{P}(d|s,\gamma)\mathcal{G}(\gamma - m_\gamma, D^{\gamma}) = \mathcal{G}\left(d-\check{R}s,N + \sum_{ab}D^\gamma_{ab}R^a ss^\dag {R^b}^\dag\right),\\
		  \check{R} &\equiv R^0 + \sum_a (m_\gamma)_a R^a.
\end{split}
\end{equation}
Compared to the result without measurements of absolute calibration, Eq.\ (\ref{likeli}), the response $R$ and the calibration covariance have been replaced by $\check{R}$ and
$D^\gamma$, respectively. This means the response became modified by new, additional, information from the absolute calibration measurements and associated the uncertainty $D^\gamma$,
which is not diagonal anymore. The resulting reference field expansion of the Hamiltonian, Eq.\ (\ref{hamgen_ref}), yields the following assignments:
\begin{equation}
\label{hamdefs_ref}
\begin{split}
\check{D}=~& \left(S^{-1} + \check{R}^\dag N^{-1}\check{R}\right)^{-1},~~~\check{j}= \check{R}^\dag N^{-1} d,~~~\check{m} = \check{D}\check{j},~~~M^{\check{~} x} \equiv \check{R}^\dag N^{-1} R^x,\\
\Lambda^{(1)} \phi =~& \frac{1}{\delta t}\underbrace{\left(\check{m}^\dag \check{D}^{-1} - \check{j}^\dag\right)}_{=0}\phi + \sum_{ab} D^\gamma_{ab}\bigg\{\check{m}^\dag M^{ab}\phi
		      - \frac{1}{2}{j^a}^\dag \left(\phi \check{m}^\dag + \check{m} \phi^\dag \right)j^b \\
		      &~ -\frac{1}{2}\check{m}^\dag M^{\check{~} a}\left(\phi \check{m}^\dag + \check{m} \phi^\dag \right)M^{b\check{~}}\check{m}
		      - \frac{1}{2}\check{m}^\dag M^{\check{~} a}\check{m} \check{m}^\dag M^{b\check{~}}\phi  +  {j^a}^\dag \check{m} \check{m}^\dag M^{b\check{~}}\phi \\
		      &~ + \frac{1}{2} {j^a}^\dag \left(\phi \check{m}^\dag + \check{m}\phi^\dag \right)M^{b\check{~}}\check{m} +  \frac{1}{2}\check{m}^\dag M^{\check{~} a} \left(\phi \check{m}^\dag + \check{m}\phi^\dag \right){j^b} \bigg\},\\	      
\Lambda^{(2)}[\phi,\phi] =&~ \frac{1}{2}\sum_{ab} D^\gamma_{ab} \bigg\{ \phi^\dag M^{ab}\phi - {j^a}^\dag \phi \phi^\dag j^b - \phi^\dag M^{\check{~} a}\check{m} \check{m}^\dag M^{b\check{~}}\phi
			   -  \check{m}^\dag M^{\check{~} a}\phi \phi^\dag M^{b\check{~}}\check{m}\\
			   &~ - \phi^\dag M^{\check{~} a}\left(\phi \check{m}^\dag + \check{m} \phi^\dag \right)M^{b\check{~}}\check{m}  - \check{m}^\dag M^{\check{~} a}\left(\phi \check{m}^\dag + \check{m} \phi^\dag \right)M^{b\check{~}}\phi \\
			   &~ + {j^a}^\dag \left(\phi \check{m}^\dag + \check{m}\phi^\dag \right)M^{b\check{~}}\phi + \phi^\dag M^{\check{~} a}\left(\phi \check{m}^\dag + \check{m}\phi^\dag \right){j^b}
			    + {j^a}^\dag \phi \phi^\dag M^{b\check{~}}\check{m} + \check{m}^\dag M^{\check{~} a}\phi \phi^\dag j^b\bigg\}\\
			   &~+ 1 ~\text{perm.},\\
\Lambda^{(3)}[\phi,\phi,\phi] =~& -  \sum_{ab} D^\gamma_{ab}\bigg\{ \frac{1}{2}\phi^\dag M^{\check{~} a}\left(\phi \check{m}^\dag + \check{m} \phi^\dag \right)M^{b\check{~}}\phi 
			       + \frac{1}{2}\check{m}^\dag M^{\check{~} a}\phi \phi^\dag M^{b\check{~}}\phi \\
			       &~ + \frac{1}{2}\phi^\dag M^{\check{~} a}\phi \phi^\dag M^{b\check{~}}\check{m}
			       - \frac{1}{2}{j^a}^\dag \phi \phi^\dag M^{b\check{~}}\phi  - \frac{1}{2}\phi^\dag M^{\check{~} a}\phi \phi^\dag j^b\bigg\} + 5~\text{perm.},\\
\Lambda^{(4)}[\phi,\phi,\phi,\phi] =~&-\frac{1}{2}~\sum_{ab}D^\gamma_{ab}\phi^\dag M^{\check{~} a} \phi \phi^\dag M^{b\check{~}} \phi + 23~\text{perm.}.
\end{split}
\end{equation}


\end{widetext}
\clearpage
\bibliography{bibliography}

\providecommand{\noopsort}[1]{}\providecommand{\singleletter}[1]{#1}%
\begin{thebibliography}{9}
\expandafter\ifx\csname natexlab\endcsname\relax\def\natexlab#1{#1}\fi
\expandafter\ifx\csname bibnamefont\endcsname\relax
  \def\bibnamefont#1{#1}\fi
\expandafter\ifx\csname bibfnamefont\endcsname\relax
  \def\bibfnamefont#1{#1}\fi
\expandafter\ifx\csname citenamefont\endcsname\relax
  \def\citenamefont#1{#1}\fi
\expandafter\ifx\csname url\endcsname\relax
  \def\url#1{\texttt{#1}}\fi
\expandafter\ifx\csname urlprefix\endcsname\relax\def\urlprefix{URL }\fi
\providecommand{\bibinfo}[2]{#2}
\providecommand{\eprint}[2][]{\url{#2}}

\bibitem[{\citenamefont{{En{\ss}lin} et~al.}(2013)\citenamefont{{En{\ss}lin},
  {Junklewitz}, {Winderling}, {Greiner}, and {Selig}}}]{2013arXiv1312.1349E}
\bibinfo{author}{\bibfnamefont{T.~A.} \bibnamefont{{En{\ss}lin}}},
  \bibinfo{author}{\bibfnamefont{H.}~\bibnamefont{{Junklewitz}}},
  \bibinfo{author}{\bibfnamefont{L.}~\bibnamefont{{Winderling}}},
  \bibinfo{author}{\bibfnamefont{M.}~\bibnamefont{{Greiner}}},
  \bibnamefont{and} \bibinfo{author}{\bibfnamefont{M.}~\bibnamefont{{Selig}}},
  \bibinfo{journal}{ArXiv e-prints}  (\bibinfo{year}{2013}),
  \eprint{1312.1349}.

\bibitem[{\citenamefont{{En{\ss}lin} et~al.}(2009)\citenamefont{{En{\ss}lin},
  {Frommert}, and {Kitaura}}}]{2009PhRvD..80j5005E}
\bibinfo{author}{\bibfnamefont{T.~A.} \bibnamefont{{En{\ss}lin}}},
  \bibinfo{author}{\bibfnamefont{M.}~\bibnamefont{{Frommert}}},
  \bibnamefont{and} \bibinfo{author}{\bibfnamefont{F.~S.}
  \bibnamefont{{Kitaura}}}, \bibinfo{journal}{\prd}
  \textbf{\bibinfo{volume}{80}}, \bibinfo{eid}{105005} (\bibinfo{year}{2009}),
  \eprint{0806.3474}.

\bibitem[{\citenamefont{Osborne}(1991)}]{1991}
\bibinfo{author}{\bibfnamefont{C.}~\bibnamefont{Osborne}},
  \bibinfo{journal}{International Statistical Review / Revue Internationale de
  Statistique} \textbf{\bibinfo{volume}{59}}, \bibinfo{pages}{pp. 309}
  (\bibinfo{year}{1991}), ISSN \bibinfo{issn}{03067734}.

\bibitem[{\citenamefont{{En{\ss}lin} and
  {Frommert}}(2011)}]{2011PhRvD..83j5014E}
\bibinfo{author}{\bibfnamefont{T.~A.} \bibnamefont{{En{\ss}lin}}}
  \bibnamefont{and}
  \bibinfo{author}{\bibfnamefont{M.}~\bibnamefont{{Frommert}}},
  \bibinfo{journal}{\prd} \textbf{\bibinfo{volume}{83}}, \bibinfo{eid}{105014}
  (\bibinfo{year}{2011}), \eprint{1002.2928}.

\bibitem[{\citenamefont{Bayes}(1763)}]{Bayes01011763}
\bibinfo{author}{\bibfnamefont{T.}~\bibnamefont{Bayes}},
  \bibinfo{journal}{Phil. Trans. of the Roy. Soc.}
  \textbf{\bibinfo{volume}{53}}, \bibinfo{pages}{370} (\bibinfo{year}{1763}).

\bibitem[{\citenamefont{Wiener}(1949)}]{wiener1964time}
\bibinfo{author}{\bibfnamefont{N.}~\bibnamefont{Wiener}},
  \emph{\bibinfo{title}{Extrapolation, Interpolation, and Smoothing of
  Stationary Time Series}} (\bibinfo{publisher}{New York: Wiley},
  \bibinfo{year}{1949}), ISBN \bibinfo{isbn}{9780262730051}.

\bibitem[{\citenamefont{{Bridle} et~al.}(2002)\citenamefont{{Bridle},
  {Crittenden}, {Melchiorri}, {Hobson}, {Kneissl}, and
  {Lasenby}}}]{2002MNRAS.335.1193B}
\bibinfo{author}{\bibfnamefont{S.~L.} \bibnamefont{{Bridle}}},
  \bibinfo{author}{\bibfnamefont{R.}~\bibnamefont{{Crittenden}}},
  \bibinfo{author}{\bibfnamefont{A.}~\bibnamefont{{Melchiorri}}},
  \bibinfo{author}{\bibfnamefont{M.~P.} \bibnamefont{{Hobson}}},
  \bibinfo{author}{\bibfnamefont{R.}~\bibnamefont{{Kneissl}}},
  \bibnamefont{and} \bibinfo{author}{\bibfnamefont{A.~N.}
  \bibnamefont{{Lasenby}}}, \bibinfo{journal}{\mnras}
  \textbf{\bibinfo{volume}{335}}, \bibinfo{pages}{1193} (\bibinfo{year}{2002}),
  \eprint{astro-ph/0112114}.

\bibitem[{\citenamefont{{Metropolis} et~al.}(1953)\citenamefont{{Metropolis},
  {Rosenbluth}, {Rosenbluth}, {Teller}, and {Teller}}}]{1953JChPh..21.1087M}
\bibinfo{author}{\bibfnamefont{N.}~\bibnamefont{{Metropolis}}},
  \bibinfo{author}{\bibfnamefont{A.~W.} \bibnamefont{{Rosenbluth}}},
  \bibinfo{author}{\bibfnamefont{M.~N.} \bibnamefont{{Rosenbluth}}},
  \bibinfo{author}{\bibfnamefont{A.~H.} \bibnamefont{{Teller}}},
  \bibnamefont{and} \bibinfo{author}{\bibfnamefont{E.}~\bibnamefont{{Teller}}},
  \bibinfo{journal}{\jcp} \textbf{\bibinfo{volume}{21}}, \bibinfo{pages}{1087}
  (\bibinfo{year}{1953}).

\bibitem[{\citenamefont{{Selig} et~al.}(2013)\citenamefont{{Selig}, {Bell},
  {Junklewitz}, {Oppermann}, {Reinecke}, {Greiner}, {Pachajoa}, and
  {En{\ss}lin}}}]{2013arXiv1301.4499S}
\bibinfo{author}{\bibfnamefont{M.}~\bibnamefont{{Selig}}},
  \bibinfo{author}{\bibfnamefont{M.~R.} \bibnamefont{{Bell}}},
  \bibinfo{author}{\bibfnamefont{H.}~\bibnamefont{{Junklewitz}}},
  \bibinfo{author}{\bibfnamefont{N.}~\bibnamefont{{Oppermann}}},
  \bibinfo{author}{\bibfnamefont{M.}~\bibnamefont{{Reinecke}}},
  \bibinfo{author}{\bibfnamefont{M.}~\bibnamefont{{Greiner}}},
  \bibinfo{author}{\bibfnamefont{C.}~\bibnamefont{{Pachajoa}}},
  \bibnamefont{and} \bibinfo{author}{\bibfnamefont{T.~A.}
  \bibnamefont{{En{\ss}lin}}}, \bibinfo{journal}{ArXiv e-prints}
  (\bibinfo{year}{2013}), \eprint{1301.4499}.

\end{thebibliography}

\end{document}